\documentclass[aps,prd,reprint,amsfonts,amssymb,amsmath,showpacs,floatfix,superscriptaddress,preprintnumbers,notitlepage,nofootinbib]{revtex4-2}

\usepackage{bm}
\usepackage{verbatim}
\usepackage{graphicx}
\usepackage{hyperref}
\usepackage{xcolor}
\usepackage{booktabs}
\usepackage{multirow}
\usepackage[margin=2.5cm]{geometry}
\usepackage{physics}

\newcommand{\epump}{\texttt{ePump}}
\newcommand{\mchess}{\texttt{mc2hessian}}

\newcommand{\xm}{x_{-}}
\newcommand{\xp}{x_{+}}
\newcommand{\xz}{x_{0}}
\newcommand{\xd}{x_{d}}
\newcommand{\msbar}{\overline{\text{MS}}}

\begin{document}
\title{\textbf{Impact of Window Observables\\ on the Transversity Quark Parton Distribution Function}}

\author{Daniel Adamiak}
\email[e-mail: ]{adamiak1@msu.edu}
\affiliation{Michigan State University, East Lansing, MI 48824, U.S.A.}
\author{Christopher J.~Monahan}
\email[e-mail: ]{cjmonahan@coloradocollege.edu}
\affiliation{Department of Physics, Colorado College, Colorado Springs, CO 80903, U.S.A.}

\date{\today}

\begin{abstract}
We study the potential impact of lattice QCD window observables on nucleon
isovector quark transversity parton distribution functions (PDFs) using the \epump{} Hessian-updating
framework. Window observables, recently proposed in
Ref.~\cite{Karpie:2025ikc} as benchmark quantities enabling direct
quantitative comparison between lattice QCD and phenomenological PDF
extractions, are weighted integrals of PDFs over a range of
Bjorken-$x$ for which both approaches are reliable. Experimental determinations of transversity PDFs carry
uncertainties of 50--100\%, while current lattice calculations have 
achieved comparable or better precision, creating an opportunity for
lattice inputs to significantly constrain a global analysis. We evaluate
the constraining power of existing lattice QCD computations of transversity
window moments and Gaussian windows against current phenomenological PDF.
We additionally
examine prospective tensor charges in the transversity channel and estimate the PDF uncertainty reduction achievable at current
and near-future lattice precision. Our results provide quantitative
guidance on which window observables will most constrain global PDF analyses.
\end{abstract}

\maketitle

\section{Introduction}
\label{sec:intro}

Parton distribution functions (PDFs) encode the longitudinal momentum
structure of the nucleon and are essential inputs to precision calculations
at hadron colliders and to efforts to understand the origin of hadron spin. PDFs arise through quantum chromodynamics (QCD) factorization of hadronic processes and are typically determined by global QCD analyses that fit large ensembles of experimental data. For
unpolarized PDFs the programme is mature~\cite{Hou:2019efy, Bailey:2020ooq,
NNPDF:2021njg, Anderson:2024evk}, but for spin-dependent distributions the situation is far less satisfactory. The transversity quark PDF $\delta q(x)$, which represents the distribution of transversely polarized quarks in a transversely polarized
nucleon, is the least known of the three leading-twist collinear
distributions. The transversity quark PDF cannot be measured in
inclusive deep-inelastic scattering (DIS), because it is chiral-odd, and must be accessed through processes involving an additional chiral-odd partner, such as di-hadron semi-inclusive DIS (SIDIS) or $e^+e^-$ annihilation into di-hadrons. The QCD factorization of these processes involves additional non-perturbative unknowns, the di-hadron fragmentation functions, and the resulting phenomenological extractions
carry uncertainties of 50--100\% across most of the range of Bjorken-$x$~\cite{Radici:2018iag, Benel:2019mcq}.

Lattice QCD offers a complementary, first-principles approach to calculating nucleon
structure. Historically, the primary lattice input to PDFs has been through
direct calculations of the lowest Mellin moments of local twist-two
operators. PDFs, however, typically cannot be reconstructed from knowledge of just the lowest Mellin moments~\cite{Detmold:2003rq,Detmold:2003tm}, but a new approach that relies on the properties of the gradient flow~\cite{Narayanan:2006rf,Luscher:2010iy,Luscher:2011bx,Luscher:2013cpa} has recently been developed to extract higher Mellin moments and provide tighter constraints on PDFs~\cite{Shindler:2023xpd,Harlander:2025qsa,Francis:2025pgf,Francis:2025rya}.\footnote{The first application of the gradient flow to lattice calculations of PDFs was proposed in a different context in \cite{Monahan:2016bvm,Monahan:2017hpu}.} 

Large-momentum effective theory (LaMET)~\cite{Ji:2013dva, Ji:2014gla} and short-distance factorization (SDF)~\cite{Radyushkin:2017cyf, Orginos:2017kos} enable lattice
calculations to access the $x$-dependence of PDFs more directly through
non-local matrix elements. PDF sets that are well-constrained by experimental data, which provide important benchmarks for the validation of newer lattice methods, whereas PDF sets that are poorly constrained by experiment offer an opportunity for lattice results to impact directly our knowledge of hadron structure. In particular, lattice results have proven particularly powerful for the quark transversity PDF, because calculations at or near the
physical pion mass have already achieved uncertainties comparable to or
smaller than current phenomenological fits~\cite{Alexandrou:2018eet,
Gao:2026wlz}. The quark transversity PDF is therefore a compelling choice to study the impact of lattice results on global analyses of PDFs. 

 A central challenge for combining lattice results with phenomenological PDF
fits is the mismatch in the domains of reliability. Global analyses are
data-driven over a limited range of $x$ and must extrapolate outside of this range, in particular at large $x$. Lattice determinations are most reliable within a range of $x$ corresponding to approximately $x\in [0.2,0.8]$, although the precise range depends on the systematic uncertainties associated with specific lattice calculations. Furthermore, full Mellin moments require integration over
all $x \in [0,1]$, including the $x$ regions where both
experimental data and lattice constraints from $x$-dependent methods are sparse. This motivates the use of observables defined
within a shared window of $x$ where both approaches are reliable. This approach has been extremely successful in the context of the hadronic vacuum polarization contribution to the muon anomalous magnetic moment~\cite{RBC:2018dos,Aliberti:2025beg}.

In Ref.~\cite{Karpie:2025ikc}, Karpie \textit{et al.}\ proposed
\emph{window observables} as benchmark quantities for quantitative
comparison between lattice QCD calculations and between lattice results and phenomenological PDF extractions.
Two examples were introduced, window moments and Gaussian
windows, and were studied using lattice data for the isovector
$u{-}d$ transversity PDF. Window observables exploit the non-trivial point-by-point correlations in the $x$-dependent results extracted from lattice data, which ensure that integrals over $x$ can be determined more
precisely than the point-wise uncertainty might suggest. These observables therefore enable meaningful comparison not just between global fits and lattice results, but also precision comparison between lattice calculations themselves.

The goal of this work is to quantify, from the phenomenological
side, how much lattice QCD window observables could reduce PDF uncertainties
if incorporated into a global analysis. A number of approaches have been used to incorporate lattice results into global analyses~\cite{Lin:2017stx,Gamberg:2022kdb,JeffersonLabAngularMomentumJAM:2022aix,Karpie:2023nyg,Avkhadiev:2025wps,Barry:2025wjx,Good:2025nny}, but here we employ the \epump{} (error PDF
Updating Method Package)~\cite{Schmidt:2018hvu, Hou:2019gfw} framework,
which uses the Hessian approximation to estimate the impact of new data
on an existing PDF set and its uncertainties at a fraction of the
computational cost of a full refit. We profile over a range of windows of different midpoints and widths. We explore their impact
 on the transversity tensor charges and isovector momentum fraction. We also explore the ability of the window and Gaussian moments to constrain the full Mellin moments. These results are compared to a profiling of the full Mellin moments to ascertain the uncertainty reduction trade-off between window, Gaussian and full Mellin moments. Lastly, we perform a double window scan to explore the effects of profiling on two window moments.

This paper is structured as follows. Section~\ref{sec:framework} reviews
the window observable formalism, the \epump{} framework, and the
Monte Carlo-to-Hessian conversion that enables \epump{} to be applied to
replica-based PDF sets. Section~\ref{sec:observables}
defines the full set of window observables we propose to test, including
both existing computations and prospective targets, and our analysis methodology and benchmarks are given in Section~\ref{sec:metrics}. We present results in Section \ref{sec:results} and summarize in Section \ref{sec:results_summary}. 

The code for our analysis can be found on GitHub, \cite{github}.

\section{Theoretical Framework}
\label{sec:framework}

\subsection{Motivation: transversity quark PDFs}
\label{sec:channels}

A critical question for this study, and for the broader structure community, is: Which PDF channels offer the
greatest opportunity for lattice-driven uncertainty reduction? The answer
is not determined solely by the lattice precision achievable, but by the
ratio of lattice precision to the existing phenomenological
uncertainty and how the information from the lattice observable is incorporated into the phenomenology. In \cite{Cocuzza:2023oam}, the Jefferson Angular Momentum collaboration (JAM) incorporated lattice observables as data; they produced Monte-Carlo replicas that resampled within the lattice data uncertainty.  One could instead choose not to resample or to include lattice data as a constraint. We consider only the case of incorporating lattice inputs as data, since it has precedent in phenomenology, appropriately propagates uncertainties,  and is amenable to Hessian updating.

The isovector transversity quark PDF $\delta u(x) - \delta d(x)$ provides a compelling example for improving the precision of phenomenological extractions through the incorporation of lattice results. First, the isovector combination benefits from the cancellation of
disconnected quark diagrams, removing one of the main sources of
systematic uncertainty in lattice calculations. Second, the transversity quark PDF is chiral-odd, so it cannot be
measured in inclusive DIS and must be extracted from SIDIS processes via
the Collins mechanism or di-hadron production, both of which require
knowledge of additional fragmentation functions. The resulting
phenomenological determinations carry relative uncertainties of
50--100\% over most of the measured $x$ range~\cite{Radici:2018iag, Benel:2019mcq}. In contrast, lattice QCD calculations using LaMET and
SDF at or near the physical pion mass have already achieved uncertainties
of $\sim$10--30\% over the range $x \in [0.1, 0.7]$~\cite{
Alexandrou:2018eet, Gao:2026wlz, HadStruc:2021qdf},
and in some comparisons the lattice result has better nominal
precision than the global phenomenological fits~\cite{Begel:2022kwp,
Burkert:2022hjz}.  Lattice calculations are therefore well-positioned to directly reduce quark transversity PDF uncertainties. Moverover, the existing window observable computations of
Ref.~\cite{Karpie:2025ikc} were performed specifically for the isovector
transversity channel.

\subsection{Window Observables}
\label{sec:windows}

The two examples of window observable introduced in Ref.~\cite{Karpie:2025ikc} are window moments and Gaussian windows. Both are defined as weighted integrals of a PDF $f(x, \mu^2)$ over a restricted interval $[\xm, \xp]$ of
Bjorken-$x$.

The window moment of order $n$ is the truncated
Mellin integral
\begin{equation}
  a_n(\xm, \xp;\, \mu^2) = \int_{\xm}^{\xp} dx\; x^n\, f(x, \mu^2)\,,
  \label{eq:window_moment}
\end{equation}
which reduces to the standard $(n{+}1)$th Mellin moment $a_{n+1}$ in the
limits $\xm \to 0$, $\xp \to 1$:
\begin{equation}
  a_n(\mu^2) = \int_{-1}^{1} dx\, x^{n-1} f(x,\mu^2)\,.
  \label{eq:mellin}
\end{equation}
For small $n$ the weight $x^n$ is broad
and the observable integrates over wide features in $x$, making it most
sensitive to the low Fourier modes of the PDF where lattice precision is
highest. As $n$ increases, the weight concentrates near $\xp$, providing
finer spatial resolution at the cost of a worse signal-to-noise ratio in
the lattice calculation.

The Gaussian window observable is defined as
\begin{align}
  {} & g_n(\xm, \xp;\, \mu^2) = \\ 
  {} & \quad \int_{-1}^{1} dx\; \sqrt{\frac{n^2}{2\pi
  \xd^2}}\; \exp\!\left[-\frac{n^2(x - \xz)^2}{2\xd^2}\right]
  f(x, \mu^2)\,, \nonumber
  \label{eq:gaussian_window}
\end{align}
where $\xz = (\xm + \xp)/2$ is the window centre and $\xd = \xp - \xm$
is its width. As $|n| \to \infty$ the Gaussian narrows and $g_n \to
f(\xz,\mu^2)$, recovering a pointwise evaluation; at $n{=}1$ the
observable probes the PDF averaged over the full window with a Gaussian
profile.

Both observables can be expressed as integrals in Ioffe time,
$\nu$, via a convolution theorem~\cite{Karpie:2025ikc}, connecting them
directly to the pseudo-ITD matrix elements computed on the lattice in the SDF framework. This dual representation allows for these quantities to be
computed on any global-analysis PDF set by direct $x$-integration, and
computed from lattice data either in $x$- or $\nu$-space. In this work, we study window observables for the CP-projected (valence) combination $q_+(x) = f(x) - f(-x)$.

\subsection{The \epump{} Framework}
\label{sec:epump}
We estimate the impact of prospective data by employing Hessian updating, also known as Hessian profiling, from 
the \epump{} package~\cite{Schmidt:2018hvu, Hou:2019gfw}. This allows us to estimate the impact of new
(pseudo-)data on an existing Hessian PDF set without performing a full
global refit, in a computationally efficient manner. In a Hessian PDF set, the uncertainty is encoded in $2N$ eigenvector error
sets $\{f_k^\pm\}$, $k = 1, \ldots, N$, together with a central best-fit
set $S_0$.

Given a new observable $O$ with measured value $O_\text{exp}$ and
uncertainty $\delta O_\text{exp}$, \epump{} minimises the combined
$\chi^2$,
\begin{equation}
  \chi^2_\text{total} = \chi^2_\text{prior}
    + \frac{(O_\text{theory}[f] - O_\text{exp})^2}{(\delta
    O_\text{exp})^2}\,,
  \label{eq:chi2}
\end{equation}
within the Hessian approximation to produce updated central and
eigenvector PDF sets. The $\chi^2_{\rm{prior}}$ is the penalty of deviating from the original fit, and is given by the quadratic sum of the normalized error sets,
\begin{equation}
    \chi^2_{\rm{prior}} = T^2 \sum\limits_iz_i^2,
\end{equation}
where $z_i$ are the coefficients of the error PDF, $(f^+_i-f^-_i)/2$, and $T^2$ is a tolerance parameter that determines the relative pull between the old fit and the new data. From the way the error sets are defined from the \mchess{} conversion, where the distribution of PDFs is normal, the appropriate choice of tolerance is $T=1$.
The fractional reduction in the uncertainty on any
observable can then be read from the updated error sets. 

As demonstrated in Ref.~\cite{Hou:2019gfw}, \epump{} reproduces full
global refit results to high accuracy for data consistent with the prior
that satisfy the Gaussian approximation. In the region where PDFs are constrained by data, this assumption has been validated, \cite{Watt:2012tq,Carrazza:2015aoa}, while at very large or small $x$, it starts to break down, \cite{Hou:2016sho, Andersen:2016qtm}. These considerations have been examined for the unpolarized PDFs and not the transversity PDFs. As such, there is no guarantee that the updated uncertainties presented in this paper are certain to be the uncertainties obtained from refit, but only approximately so. This is acceptable for our purposes, because we are investigating the relative impact of the window, Gaussian, and full Mellin moments, based on reasonable choices of relative uncertainty that can be expected from future lattice calculations, rather than assessing the impact of specific, known lattice results. The relative impact should still be preserved, even if the Gaussian approximation is violated slightly. The other key assumption behind Hessian updating for uncertainty reduction quantification is that the new data are not too far away from the data of the old fit. In the case of 'incompatible data', even precise new data won't reduce the overall uncertainty very much, because it is unclear which data are correct. While this case is interesting--we would want to know if the predictions of lattice QCD disagree with experiment--this can only be investigated with lattice results that contain a complete, quantitative error budget. What we investigate in this paper is a prescription for prioritizing the computation of specific lattice observables, ordered by the impact on the uncertainties assuming perfect agreement between lattice and experiment. As such, the lattice observables, the window, Gaussian and Mellin moments, are computed assuming the central value of the existing PDFs is correct.

\subsection{Monte Carlo to Hessian Conversion}
\label{sec:mc2hess}

A prerequisite for applying \epump{} is a Hessian representation of the
baseline PDF set. For transversity distributions, the major global analyses
use Monte Carlo replica methods. In particular, the JAM results~\cite{Cammarota:2020qcw,
Cocuzza:2023oam} and
the analyses of Radici and Bacchetta~\cite{Radici:2018iag} do not provide results directly in Hessian form.

We convert these replica sets to Hessian eigenvector representations
using the \mchess{} algorithm~\cite{Carrazza:2015aoa}, which selects an
optimal linear basis from a subset of the Monte Carlo replicas using a
genetic algorithm. The methodology faithfully reproduces the variance,
covariance, and non-Gaussian tails of the original replica ensemble with
minimal information loss, and has been validated against native Hessian
sets~\cite{Carrazza:2015aoa}. An independent implementation by the CT
group~\cite{Hou:2016sho} confirms the robustness of this approach. With
the resulting Hessian sets in hand, \epump{} can be applied to polarized
PDFs in precisely the same way as to unpolarized ones.

\section{Proposed Window Observables}
\label{sec:observables}

We divide this study of window observables into two categories: those
already computed from lattice QCD data, which we take directly as input
to \epump{}, and prospective observables for which we estimate anticipated
precision based on the current state of lattice calculations.

\subsection{Existing Lattice QCD Transversity Results}
\label{sec:existing}

The window observable computations of Ref.~\cite{Karpie:2025ikc} provide
the primary inspiration for this study. The authors computed window moments
$a_n(\xm,\xp)$ and Gaussian windows $g_n(\xm,\xp)$ for the isovector
$u{-}d$ transversity distribution from lattice QCD pseudo-ITD data on a
single ensemble with pion mass $m_\pi = 358\,\text{MeV}$ and lattice
spacing $a = 0.094\,\text{fm}$, using Gaussian Process Regression (GPR)
for the $x$-reconstruction. The window boundaries were fixed at $\xm =
0.1$, $\xp = 0.5$ at renormalization scale $\mu = 2\,\text{GeV}$ in the
$\msbar$ scheme.

\subsubsection*{Central values and the \texorpdfstring{$g_T$}{gT}
normalization}

A practical complication for the \epump{} analysis is that
Ref.~\cite{Karpie:2025ikc} presents the lattice QCD results for the
window observables normalized by the tensor charge $g_T = a_0(0,1)
= \int_0^1 dx\,\delta(u-d)(x)$, i.e.\ as $a_n / g_T$ and $g_n / g_T$.
This normalization is a consequence of using the ratio method to renormalise the relevant lattice operator by constructing an RGI ratio of pseudo-ITDs. This also removes correlated fluctuations in the ratio, but absorbs the overall
normalization of the tensor charge. The absolute values of the window observables are recovered by multiplying the normalized results by the $g_T$ value
obtained from a separate non-perturbative renormalization
procedure~\cite{Karpie:2025ikc}. The results for the absolute (unnormalized)
window observables are presented in Fig.~4 of Ref.~\cite{Karpie:2025ikc}
for each Wilson-line separation $z/a \in [1, 8]$, but are
not tabulated with explicit numerical values.

Furthermore, the lattice results exhibit a residual $z$-dependence: the
window observables show a mild upward trend with increasing $z$,
particularly visible in the Gaussian windows, reflecting unresolved
systematic effects that are likely generated by higher-twist contamination and discretization effects. The statistical uncertainties decrease with
increasing $z$ because the larger Ioffe-time range constrains the GPR
reconstruction. The authors of
Ref.~\cite{Karpie:2025ikc} note that a multi-ensemble study is required
to disentangle these effects.

For the \epump{} analysis, we therefore adopt the following strategy.
We take as our central values the window observables evaluated
directly from phenomenological quark transversity PDFs, specifically
the JAMDiFF fits used as reference distributions in
Ref.~\cite{Karpie:2025ikc}, integrated over $x \in [\xm, \xp]$:
\begin{equation}
  a_n^\text{pheno}(\xm, \xp) = \int_{\xm}^{\xp} dx\, x^n\,
  \delta(u - d)^\text{pheno}(x)\,.
  \label{eq:pheno_central}
\end{equation}
This choice is appropriate for the purpose of estimating the constraining
power of the lattice measurement, but future lattice calculations will enable more refined window observable inputs for careful control of all systematic effects. For this study, and in particular, for the \epump{} framework, what is important is that the lattice observable is assigned a central value that lies within the phenomenological uncertainty band, with an assigned error that reflects the lattice precision. If the lattice result is consistent with
phenomenology (as Fig.~4 of Ref.~\cite{Karpie:2025ikc} suggests at the
lowest-$n$ observables), this assignment is self-consistent. Where tension
exists, the shift in the \epump{} central PDF will reflect the pull of
the lattice data, which is itself physically interesting.

For the assigned lattice uncertainties, we take inspiration from the total errors (combining
statistical and GPR reconstruction) from Fig.~4 of
Ref.~\cite{Karpie:2025ikc} at the shortest Wilson-line separation $z/a
= 2$, where higher-twist contamination is expected to be smallest and
the results are most directly comparable to the leading-twist factorization
used in the matching. We then add in quadrature an additional
systematic uncertainty estimated from half the spread of the results
across $z/a \in [2, 6]$, as a conservative estimate of the uncontrolled
matching and discretization systematics. The resulting estimates are
given in Table~\ref{tab:existing}. From this table, we take $10\%$ to be a reasonable estimate for the relative uncertainty for prospective lattice observables. We use this estimate instead of the actual extracted value, because we are going to be scanning over window moments that have note yet been computed, whose uncertainties are unknown. By keeping relative uncertainty fixed, we are able to estimate the impact on PDF uncertainty reduction that comes only from the choice of lattice input. 

\begin{table*}[ht]
\centering
\label{tab:existing}
\setlength{\tabcolsep}{5pt}
\begin{tabular}{lcccc}
\toprule
Observable & $n$ & Central value (pheno.) & Lattice unc.\ (assigned) &
  Pheno.\ unc.\ \\
\midrule
$a_0(0.1,\,0.5)$ & 0 & $\sim\int_{0.1}^{0.5}\delta(u{-}d)\,dx$
  & $\sim 10\%$ & $\sim 40\%$ \\
$a_1(0.1,\,0.5)$ & 1 & $\sim\int_{0.1}^{0.5}x\,\delta(u{-}d)\,dx$
  & $\sim 15\%$ & $\sim 50\%$ \\
$a_2(0.1,\,0.5)$ & 2 & $\sim\int_{0.1}^{0.5}x^2\,\delta(u{-}d)\,dx$
  & $\sim 20\%$ & $\sim 60\%$ \\
$a_3(0.1,\,0.5)$ & 3 & $\sim\int_{0.1}^{0.5}x^3\,\delta(u{-}d)\,dx$
  & $\sim 30\%$ & $\sim 80\%$ \\
$g_1(0.1,\,0.5)$ & 1 & $\sim g_1^\text{JAM}$
  & $\sim 12\%$ & $\sim 40\%$ \\
$g_2(0.1,\,0.5)$ & 2 & $\sim g_2^\text{JAM}$
  & $\sim 18\%$ & $\sim 55\%$ \\
$g_3(0.1,\,0.5)$ & 3 & $\sim g_3^\text{JAM}$
  & $\sim 28\%$ & $\sim 70\%$ \\
\bottomrule
\end{tabular}
\caption{Window observables from the lattice QCD study of
Ref.~\cite{Karpie:2025ikc}, used as input pseudo-data for the primary
\epump{} analysis. All observables are for the isovector $u{-}d$
transversity at $\mu = 2\,\text{GeV}$ ($\msbar$), with window boundaries
$\xm = 0.1$, $\xp = 0.5$. Central values are taken from a direct
integration of the JAM3D or JAMDiFF quark transversity PDFs over $[\xm,\xp]$
(see text). The assigned lattice uncertainties are estimated from
Fig.~4 of Ref.~\cite{Karpie:2025ikc} as described in the text;
they include statistical, GPR reconstruction, and a conservative
$z$-dependence systematic. The phenomenological uncertainties quoted in
the final column are from the JAM3D and JAMDiFF 1$\sigma$ bands
integrated over the same window.}
\end{table*}

We note that the central values in Table~\ref{tab:existing} are left in
symbolic form because the precise numerical values depend on the
quark transversity PDF set that is used as the reference (JAM3D vs.\ JAMDiFF give
somewhat different central values). These values can be
computed directly from the published PDF sets by numerical integration. We 
compute them explicitly as part of the \epump{} implementation, and report
the numerical values in the results section.

A key point is that, even with the assigned 10--30\% lattice uncertainties,
the impact on the global transversity fit is expected to be dramatic. The
phenomenological uncertainties on $\delta(u-d)(x)$ in this $x$ range
are of order 40--80\% (Table~\ref{tab:existing}), so even the most
uncertain window observable listed carries two to three times better precision
than the corresponding phenomenological constraint. This asymmetry is
what makes the transversity channel so promising, and we expect the
\epump{} analysis to show substantial uncertainty reduction across the
quark transversity PDF eigenvectors sensitive to the $x \in [0.1, 0.5]$
region.

\subsection{Prospective Transversity Observables}
\label{sec:prosp_transversity}
With our computational pipeline, we are able to profile many lattice observables with negligible computational cost. Therefore we scan over a range of window mid-points and widths that could be computable on the lattice in the near future, computing both the window and Gaussian moments of $\delta q^{u-d}$. To provide realistic estimates of potential uncertainties on future calculations of window observables, we note that current lattice calculations with proton momenta $P_z \sim 2$--$3\,
\text{GeV}$ on ensembles at physical pion mass~\cite{Alexandrou:2018eet,
Gao:2026wlz} suggest that the $n{=}0$ and $n{=}1$ window
moments for this window can be obtained at the 15--20\% level with current
resources and that it is reasonable to expect that this will improve to $\sim$5\% with next-generation ensembles and larger hadron momenta.

We examine the impact of such profiling on a few quantities of phenomenological interest. First, the tensor charges,
\begin{eqnarray}
    g_T^{u-d} &=& \int_0^1 dx \left[\delta u(x) - \delta d(x)\right]\\
    g_T^{u} &=& \int_0^1 dx\, \delta u(x) \\ \qquad g_T^{d} &=& \int_0^1 dx\, \delta d(x),
\end{eqnarray}
which are important for BSM searches, \cite{Bhattacharya:2011qm, Courtoy:2015haa, Bhattacharya:2016zcn}, and determining the neutron electric dipole moment, \cite{Bhattacharya:2015esa}.

We also examine the isovector momentum fraction
\begin{align}
    {} & \langle x \rangle_{\delta u^+ - \delta d^+} = \\
    {} & \qquad \int_0^1 dx\, x\left[\delta u(x) + \delta\bar{u}(x) - \delta d(x) - \delta\bar{d}(x)\right],\nonumber 
\end{align}
which is a quantity frequently used as a point of comparison between phenomenological extractions of transversity PDFs and lattice computations, \cite{Mondal:2026zok}.

We then compute Hessian updating of the full Mellin moments to get a picture of how much information on the full moment is contained within the partial moments.

We also perform the same updating, but with the full Mellin moment as an input, so that we can compare the reduction in uncertainties with the two approaches.

Last, after determining which window provides the best overall improvement on the tensor charges, we compute the accumulated moments. That is, we compute an update to the PDF uncertainties the first $n$ moments are included. We compare the results from including the first $n$  window, Gaussian, and full Mellin moments. To illustrate current lattice precision for moments beyond $n=3$, albeit for the computationally simpler case of the pion, we list results from Ref.~\cite{Francis:2025rya,Francis:2025pgf} in table \ref{xnsigma}. One can expect that similar moments for the nucleon will be statistically noisier and with more significant systematic uncertainties associated with excited state contamination. Nevertheless these results provide a reasonable starting point for our impact study of the higher moments. As table \ref{xnsigma} demonstrates, higher moments are typically more uncertain, because of the increasing statistical uncertainty associated with higher derivative operators, but for simplicity and clarity we examine the impact of higher moments with fixed relative uncertainties. More realistic future benchmarking studies would relax this assumption.
\begin{table*}[ht]
\centering
\setlength{\tabcolsep}{4pt}
\renewcommand{\arraystretch}{1.2}
\begin{tabular}{cccc}
\toprule
Moment ($n$) & Central value & Uncertainty & Rel.~uncertainty (\%) \\
\midrule
3 & 0.433 & 0.014 & 3.2\\
4 & 0.275 & 0.023 & 8.4 \\
5 & 0.130 & 0.043 & 33 \\
6 & 0.158 & 0.039 & 25 \\
\bottomrule
\end{tabular}
\caption{Moments and uncertainties from  \cite{Francis:2025rya,Francis:2025pgf} for the pion. \label{xnsigma}}
\end{table*}
\begin{table*}[ht]
\centering
\label{tab:lattice_errors}
\setlength{\tabcolsep}{4pt}
\renewcommand{\arraystretch}{1.2}
\begin{tabular}{ccccccc}
\toprule
\multirow{2}{*}{Observable} & \multicolumn{3}{c}{$N_f = 2+1+1$}  & \multicolumn{3}{c}{$N_f = 2+1$} \\
& Value & Rel.~err.~(\%) & Ref. & Value & Rel.~err.~(\%) & Ref.\\
\midrule
$g^{u-d}_T$ &  0.981(21) & 2.1 & \cite{Jang:2023zts,Alexandrou:2022dtc} & 0.993(15) & 1.5 & \cite{Park:2021ypf,QCDSFUKQCDCSSM:2023qlx,Bali:2023sdi,Djukanovic:2024krw} \\
\\
$g_T^u$ & 0.784(28)(10) & 3.8 & \cite{Gupta:2018lvp} & \quad --- & \quad --- & \quad ---  \\
$g_T^d$ & \!\!\!\!\!$-0.204(11)(10)$ & 7.3 & \cite{Gupta:2018lvp} & \quad --- & \quad --- & \quad --- \\
$g_T^s$ & \!\!\!\!\!$-0.0027(16)$ & 59.3 & \cite{Gupta:2018lvp} & \quad --- & \quad --- & \quad --- \\
\\
$\langle x\rangle_{\delta u^+-\delta d^+}$ & 0.195(25) & 12.8 & \cite{Mondal:2020cmt,Alexandrou:2022dtc} & 0.206(17) & 8.3 &  \cite{Djukanovic:2024krw,Mondal:2020ela} \\
\bottomrule
\end{tabular}
\caption{Representative uncertainties in contemporary lattice results for first and second moments, from Ref.~\cite{FlavourLatticeAveragingGroupFLAG:2024oxs}. All renormalized quantities are expressed at a scale of 2~GeV in the $\overline{\mathrm{MS}}$-scheme. Results from a single lattice calculations are referred to as ``FLAG values'', rather than averages. These results are determined directly from lattice calculations of hadronic matrix elements of local operators. }
\end{table*}

\section{Analysis Methodology and Benchmarks}
\label{sec:metrics}

\subsection{Analytical expectation for the uncertainty ratio}
\label{sec:analytic}

Before we present our numerical results, we derive the
expected uncertainty reduction analytically for the simple case of a
single new data point. This provides a baseline against which the
numerical \epump{} output can be checked, and builds intuition for
interpreting the heatmap scans that we present in Section~\ref{sec:results}.

Suppose the new lattice datum is a window observable $X \equiv g_n(\xz,
w)$ or $a_n(\xz,w)$, with experimental (lattice) central value $X^E$ and
uncertainty $\sigma_\text{lat}$. Within the \epump{}
framework~\cite{Schmidt:2018hvu}, the key quantities are the vector of
gradient components,
\begin{equation}
  \Delta X^i = \frac{X(f^{+i}) - X(f^{-i})}{2}\,,
  \label{eq:gradient}
\end{equation}
and the rank-1 matrix $M^{ij} = \Delta X^i \Delta X^j / (T^2
\sigma_\text{lat}^2)$, which has a single non-zero eigenvalue
\begin{equation}
  \lambda^{(1)} = \frac{(\Delta X)^2}{T^2 \sigma_\text{lat}^2}\,,
  \label{eq:eigenvalue}
\end{equation}
where $\Delta X = \sqrt{\sum_i (\Delta X^i)^2}$ is the prior PDF
uncertainty on $X$.

The shift in the best-fit parameters is proportional to
\begin{equation}
A^i = (X^E - X(\mathbf{0})) \Delta X^i / (T^2 \sigma_\text{lat}^2)\,.
\end{equation}
When the lattice central value agrees with the PDF prediction,
$X^E = X(\mathbf{0})$, we have $A^i = 0$ exactly and the central PDF
does not move. This result provides a useful cross-check of the \epump{} output.

From Eq.~(30) of Ref.~\cite{Schmidt:2018hvu}, the updated symmetric
uncertainty on any observable $Y$ is
\begin{equation}
  (\Delta Y)^2_\text{new}
  = \sum_{i,j} \Delta Y^i \,(\delta + M)^{-1}_{ij}\, \Delta Y^j\,.
  \label{eq:updated_unc}
\end{equation}
Since $M^{ij}$ is rank-1, the matrix $(\delta + M)^{-1}$ only modifies
the component of $\Delta Y^i$ parallel to $\Delta X^i$. Decomposing
$\Delta Y^i$ into parallel and perpendicular components and defining the
correlation cosine
\begin{equation}
  \cos\theta_{XY} = \frac{\sum_i \Delta X^i \Delta Y^i}{|\Delta X|\,|\Delta Y|}\,,
  \label{eq:costheta}
\end{equation}
one obtains the uncertainty ratio
\begin{equation}
  \frac{\sigma_\text{after}}{\sigma_\text{before}}
  = \sqrt{1 - \frac{\cos^2\theta_{XY}}{1 + T^2\sigma_\text{lat}^2/(\Delta X)^2}}\,.
  \label{eq:sigma_ratio}
\end{equation}
Two limiting cases provide important checks of this general result. When $X = Y$ (so that the input and output
are identical), $\cos\theta = 1$ and Eq.~\eqref{eq:sigma_ratio} reduces
to the standard result from simple error propagation,
\begin{equation}
  \frac{1}{(\Delta X)^2_\text{new}}
  = \frac{1}{(\Delta X)^2} + \frac{1}{T^2 \sigma_\text{lat}^2}\,.
  \label{eq:harmonic}
\end{equation}
When $Y$ is orthogonal to $X$ in PDF parameter space
($\cos\theta = 0$), the uncertainty on $Y$ is unaffected regardless of the precision of the additional lattice data. The general case interpolates between these extremes,
with $\cos\theta_{XY}$ governing how efficiently the lattice constraint
on $X$ propagates to $Y$. Since $M^{ij}$ is rank-1, exactly $N-1$
eigenvector directions are completely unmodified by a single data point.

\subsection{Setup and impact metrics}
\label{sec:impact_metrics}

Our results use the JAMDiFF23 quark transversity PDF
set~\cite{Cocuzza:2023oam}, converted from Monte Carlo replica form to a
Hessian eigenvector representation via \mchess{}~\cite{Carrazza:2015aoa}. The pseudo-data
central value is computed by integrating the JAMDiFF23 central PDF member
over the window, so that $X^E = X(\mathbf{0})$ and the central PDF does
not move. The assigned lattice uncertainty is a fixed relative precision
$\varepsilon$, i.e.\ $\sigma_\text{lat} = \varepsilon\,|X(\mathbf{0})|$.
All results are at $Q^2 = 4.0\,\text{GeV}^2$.

We present $\sigma_\text{after}/\sigma_\text{before}$ as a function of
window midpoint $\xz$ and width $w$ for the following input observables:
\begin{enumerate}
    \item $a_0(\xz,w)$ (Gaussian window, $n{=}1$) at $\varepsilon = 10\%$ and
$\varepsilon = 25\%$;
    \item $a_0(\xz,w)$ (window moment, $n{=}0$) at
$\varepsilon = 10\%$; and
    \item $a_1(\xz,w)$ (window moment, $n{=}1$) at
$\varepsilon = 10\%$.
\end{enumerate}
For each input we evaluate the impact on:
\begin{enumerate}
    \item the full
Mellin moments $a_n(0,1)$ for $n = 0,1,2,3$;
    \item the tensor charges
$g_T^{u-d}$, $g_T^u$, $g_T^d$;
    \item the isovector momentum fraction
$\langle x(\delta u^+ - \delta d^+)\rangle$; and
    \item as a cross-check, the input observable itself.
    \item accumulated moments; computing and updating on multiple moments simultaneously
    \item a second scan to find which pair of window observables is optimal in constraining the tensor charges
\end{enumerate}

At this point, we highlight some details concerning our choice PDF set that affect this analysis. JAM imposes the Soffer bound in their analysis, which heavily constrains the large $x$ behavior of the transversity PDFs. As such, the impact from the Hessian updating in this work can be seen as an underestimate of the true impact. Additionally, the JAMDiFF23 replicas come from an analysis that simultaneously fits the transversity PDFs and the fragmentation functions, inducing a correlation between the two. Here, since we only update the PDFs, we ignore this implicit correlation, potentially leading to an overestimate of the impact of PDF uncertainty reduction. The relative impact from scanning over different windows should still be meaningful for deciding which lattice moments are most useful to compute. In the \mchess conversion, a number of replicas are discarded for violating the Gaussian assumption necessary for the Hessian analysis. We do not foresee this having a large impact on the qualitative results of this paper, as there does not appear to be a systematic bias in the non-Gaussianity of the replicas, except, potentially, at very large $x$, where the Soffer bound has the largest impact. 
\section{Results}
\label{sec:results}

\subsection{Mellin Moments}

To setup a baseline against which the window and Gaussian moments can be compared, we profile against the $n_{th}$ Mellin moments. We assume a $10\%$ relative uncertainty on the computation. The results for the updated Mellin moments are shown in table ~\ref{tab:mellin_moments_out}, and the updated tensor charges are in table ~\ref{tab:mellin_charges_out}.
\begin{table*}[ht]
\centering
\setlength{\tabcolsep}{8pt}
\renewcommand{\arraystretch}{1.3}
\begin{tabular*}{\textwidth}{@{\extracolsep{\fill}}c|cccc}
\toprule
$n_\text{in}$
  & $a_0^\text{out}$ (38.9\%)
  & $a_1^\text{out}$ (24.3\%)
  & $a_2^\text{out}$ (30.6\%)
  & $a_3^\text{out}$ (41.7\%) \\
\midrule
0 & \textbf{0.25} & 0.69 & 0.91 & 0.96 \\
1 & 0.73 & \textbf{0.38} & 0.57 & 0.70 \\
2 & 0.91 & 0.53 & \textbf{0.31} & 0.38 \\
3 & 0.96 & 0.65 & 0.32 & \textbf{0.23} \\
4 & 0.97 & 0.72 & 0.40 & 0.23 \\
5 & 0.98 & 0.78 & 0.49 & 0.32 \\
\bottomrule
\end{tabular*}
\caption{$\sigma_\text{after}/\sigma_\text{before}$ on the full Mellin
moments $a_n^\text{out}(0,1)$ after profiling with the full Mellin moment
$a_{n_\text{in}}(0,1)$ at $\varepsilon = 10\%$ relative precision.
Prior relative uncertainties are quoted in parentheses. Diagonal entries
(bold) are the trivial self-observable case, reproducible by simple error
propagation (Eq.~\eqref{eq:harmonic}).
JAMDiFF23, $Q^2 = 4.0\,\text{GeV}^2$.\label{tab:mellin_moments_out}}
\end{table*}

\begin{table*}[ht]
\centering
\setlength{\tabcolsep}{8pt}
\renewcommand{\arraystretch}{1.3}
\begin{tabular*}{\textwidth}{@{\extracolsep{\fill}}c|cccc}
\toprule
$n_\text{in}$
  & $g_T^{u-d}$ (38.9\%)
  & $g_T^{u}$ (20.8\%)
  & $g_T^{d}$ (794.9\%)
  & $\langle x_{\delta u-\delta d}\rangle$ (31.4\%) \\
\midrule
0 & 0.25 & 0.78 & 0.55 & 0.55 \\
1 & 0.73 & 0.89 & 0.81 & 0.51 \\
2 & 0.91 & 0.93 & 0.96 & 0.72 \\
3 & 0.96 & 0.94 & 0.99 & 0.81 \\
4 & 0.97 & 0.94 & 1.00 & 0.86 \\
5 & 0.98 & 0.95 & 1.00 & 0.88 \\
\bottomrule
\end{tabular*}
\caption{$\sigma_\text{after}/\sigma_\text{before}$ on tensor charges and
the isovector momentum fraction after profiling with the full Mellin moment
$a_{n_\text{in}}(0,1)$ at $\varepsilon = 10\%$ relative precision. Prior
relative uncertainties are quoted in parentheses. Since $g_T^{u-d} =
a_0(0,1)$, its column is identical to the $a_0^\text{out}$ column of
Table~\ref{tab:mellin_moments_out}.
JAMDiFF23, $Q^2 = 4.0\,\text{GeV}^2$.}\label{tab:mellin_charges_out}
\end{table*}

\subsection{Window and Gaussian moments: impact comparison}
\label{sec:results_comparison}

\begin{figure*}[t]
  \centering
  \includegraphics[width=\textwidth]{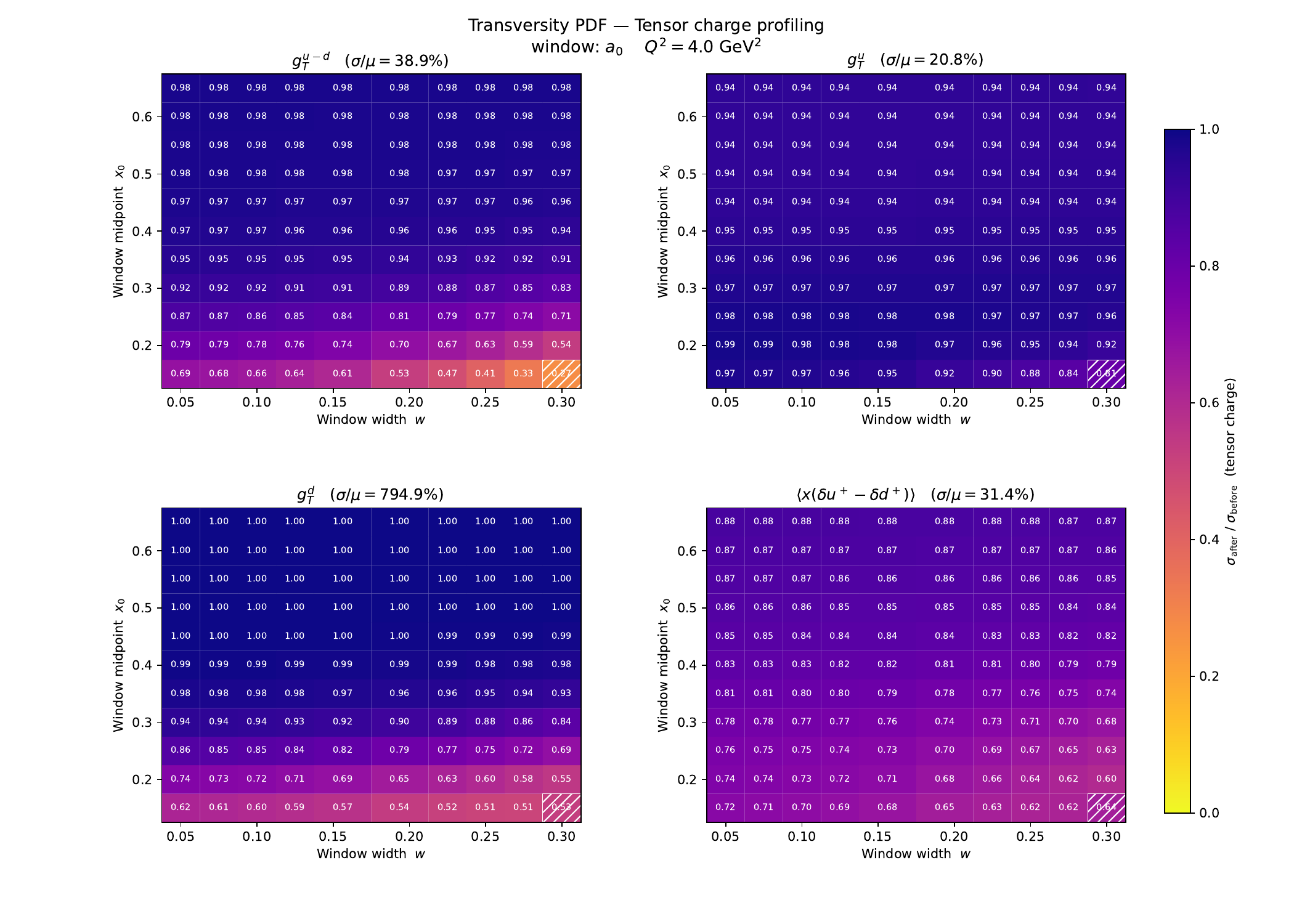}
  \caption{$\sigma_\text{after}/\sigma_\text{before}$ on tensor charges
  and the isovector momentum fraction after profiling with the flat
  window moment $a_0(\xz,w)$ at $\varepsilon = 10\%$. The four panels
  show (left to right, top to bottom) $g_T^{u-d}$, $g_T^u$, $g_T^d$,
  and $\langle x(\delta u^+ - \delta d^+)\rangle$. Prior relative
  uncertainties from JAMDiFF23 are quoted in each panel title. Hatched
  cells indicate windows clipping against $x = 0$ or $x = 1$.
  JAMDiFF23 isovector transversity, $Q^2 = 4.0\,\text{GeV}^2$.}\label{fig:charges_a0}
\end{figure*}

\begin{figure*}[t]
  \centering
  \includegraphics[width=\textwidth]{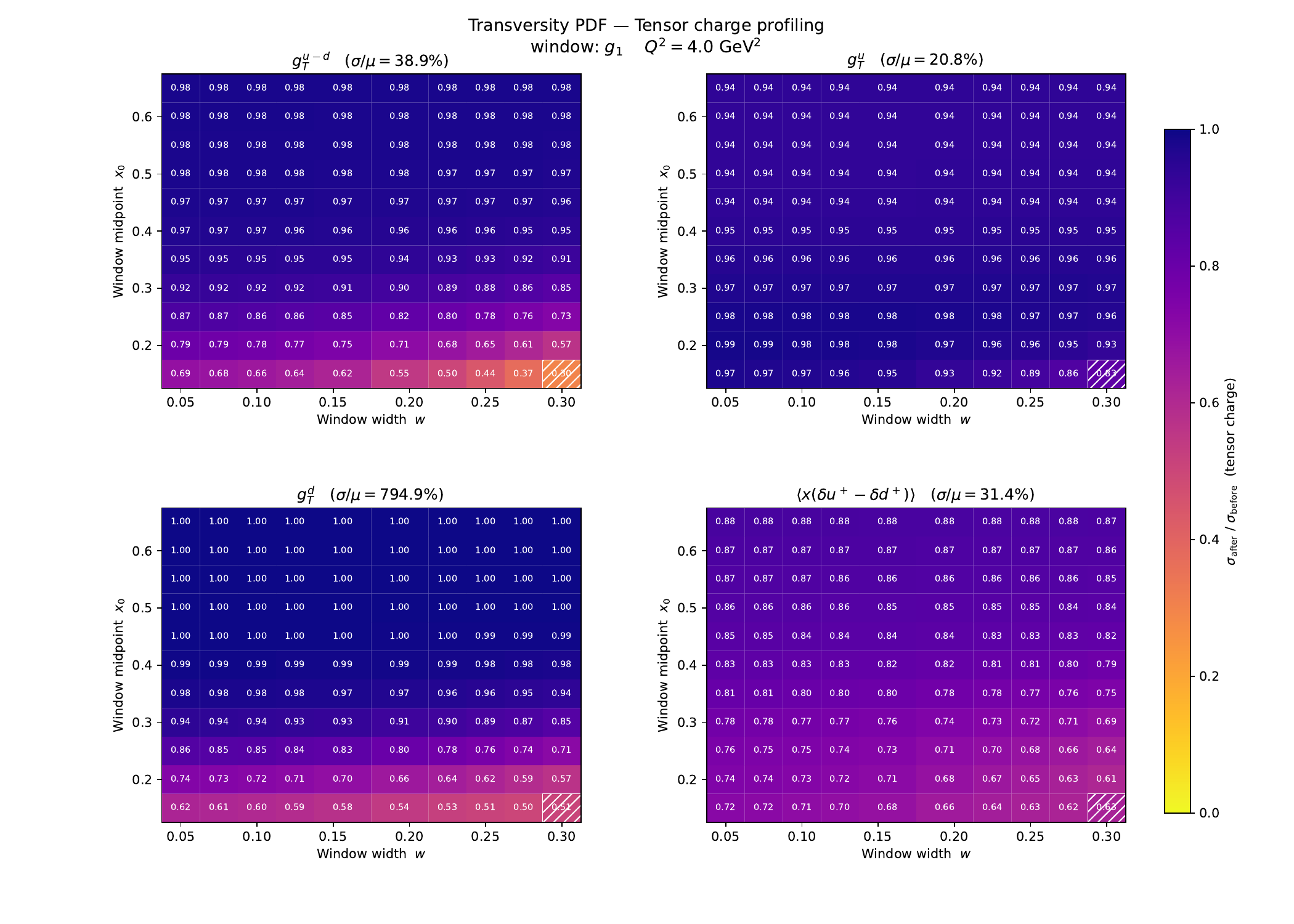}
  \caption{Same as Fig.~\ref{fig:charges_a0}, but for the Gaussian
  window $g_1(\xz,w)$ at $\varepsilon = 10\%$.}\label{fig:charges_g1}
\end{figure*}

Figure~\ref{fig:charges_a0}
shows $\sigma_\text{after}/\sigma_\text{before}$ on the tensor
charges $g_T^{u-d}, g_T^{u},g_T^{d}$, and the isovector momentum fraction for the window moment, $a_0$, at $\varepsilon = 10\%$. Figure~\ref{fig:charges_g1} show the same for the Gaussian moment, $g_1$. 

The two types of window observable are similar in their
constraining power. For the isovector tensor charge $g_T^{u-d}$, the
optimal choice in the $a_0$ scan achieves
$\sigma_\text{after}/\sigma_\text{before} \approx 0.27$ (at low $\xz$,
large $w$), compared to $\approx 0.30$ for $g_1$ at the same location, corresponding to a difference less than 0.03 across the $(\xz, w)$ plane. A
similar picture holds for all other output observables. The practical
conclusion is that the choice between window and Gaussian moments
has negligible impact on the PDF uncertainty reduction achievable at
fixed lattice precision; the window midpoint $\xz$ is far more
important than the functional form of the window weight. This conclusion is to be expected, because both the window moment and the Gaussian window carry much the same genuinely nonperturbative information from any given lattice calculation.

Both inputs show the same qualitative behavior in the flavor
decomposition. The $g_T^u$ panel is uniformly near 0.94 regardless of
window parameters, confirming that the isovector lattice calculation
cannot constrain $g_T^u$ in JAMDiFF23. The $g_T^d$ panel mirrors
$g_T^{u-d}$ in pattern and magnitude, a consequence of the algebraic
identity $g_T^d = g_T^u - g_T^{u-d}$ combined with the fact that
$g_T^u$ is uncorrelated with the window observable. The 795\% prior
uncertainty on $g_T^d$ means that even a modest reduction in
$g_T^{u-d}$ translates into a large absolute improvement in $g_T^d$.
The $\langle x(\delta u^+ - \delta d^+)\rangle$ panel shows uniformly
strong reduction ($\sigma_\text{after}/\sigma_\text{before} \approx
0.60$--$0.87$) with relatively weak $\xz$ dependence, reflecting the
good overlap between $x$-weighted integrals and both choices of window observables.

A notable feature of both heatmaps is the behavior near $\xz \approx
0.25$. The contours show a non-monotonic shoulder, with cells at $\xz
\approx 0.25$ achieving better reduction than both lower and higher
midpoints at narrow widths. This reflects a local maximum in $\cos\theta$
between the window observable and $g_T^{u-d}$, associated with the
region where the JAMDiFF23 quark transversity PDF eigenvectors have greatest
sensitivity to the isovector combination.

\subsection{Impact of lattice precision}
\label{sec:results_precision}
\begin{figure*}[h]
  \centering
  \includegraphics[width=\textwidth]{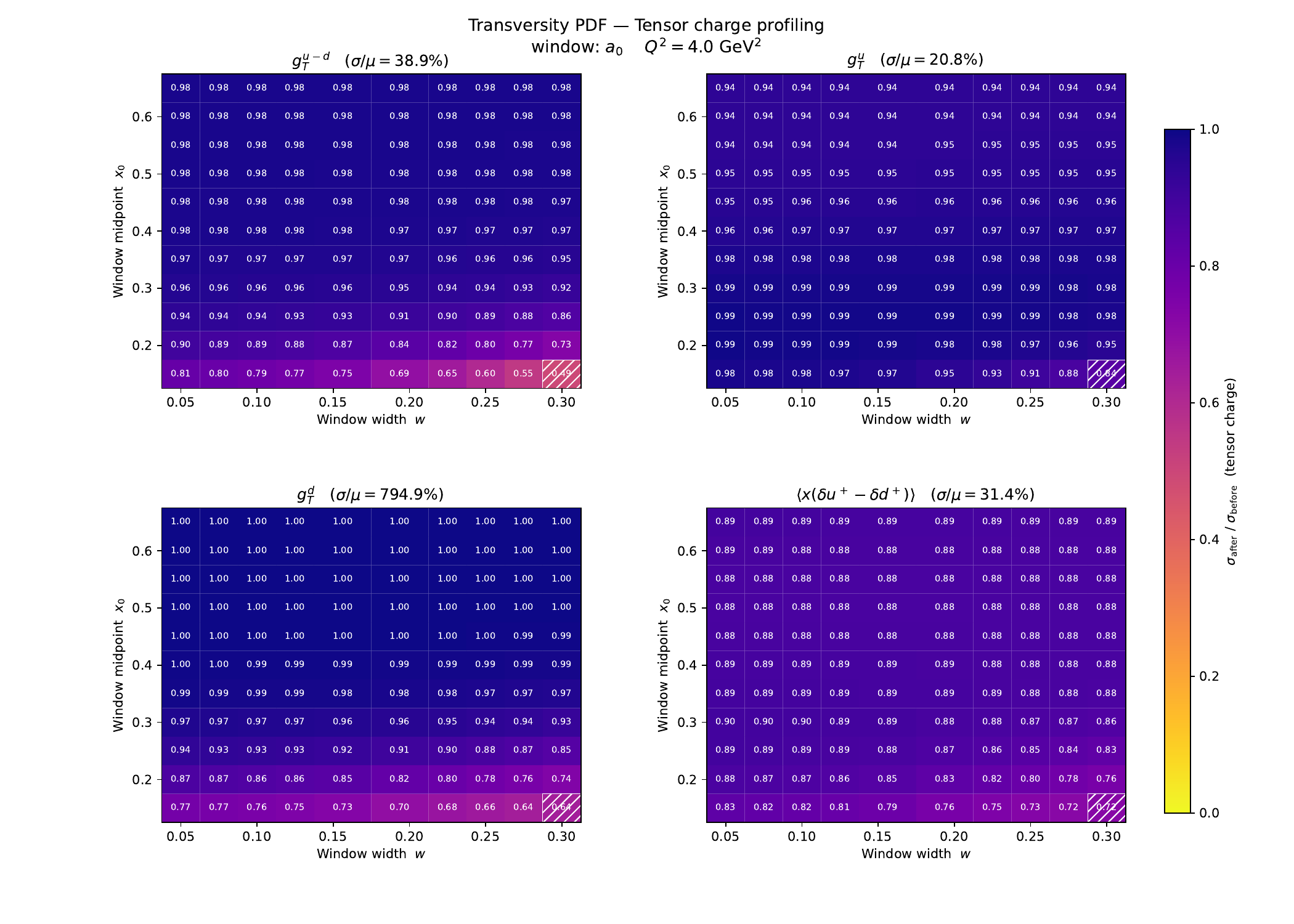}
  \caption{$\sigma_\text{after}/\sigma_\text{before}$ on the tensor charges after profiling with $a_0(\xz,w)$
  at $\varepsilon = 25\%$. Compare with the $\varepsilon = 10\%$ case
  in Fig.~\ref{fig:charges_a0}. The degradation in
  constraining power is substantial across all moments; for $g_T^{u-d}$
  ($n{=}0$) the best achievable ratio rises from $\approx 0.37$ to
  $\approx 0.55$.
  JAMDiFF23 isovector transversity, $Q^2 = 4.0\,\text{GeV}^2$.}
  \label{fig:precision_comparison}
\end{figure*}

In Figure~\ref{fig:precision_comparison} we show $a_0(\xz,w)$ at
$\varepsilon = 25\%$, which should be compared to Figure~\ref{fig:charges_a0}, which shows  $a_0(\xz,w)$ at
$\varepsilon = 10\%$ for the tensor charges.

The degradation in constraining power is substantial: for $g_T^{u-d}$
(the $n{=}0$ panel), the best achievable
$\sigma_\text{after}/\sigma_\text{before}$ rises from $\approx 0.37$ at
$\varepsilon = 10\%$ to $\approx 0.55$ at $\varepsilon = 25\%$, confirming
that lattice precision is the dominant constraint on potential improvements.
A factor of 2.5 in relative precision translates to roughly a factor of
two in uncertainty reduction on the tensor charge.

The reduction in constraining power is more modest for the other tensor charges, which show an absolute difference of 10$\%$ in the best candidate windows.

\subsection{Impact of order on Window moments}
\begin{figure*}[t]
  \centering
  \includegraphics[width=1\textwidth]{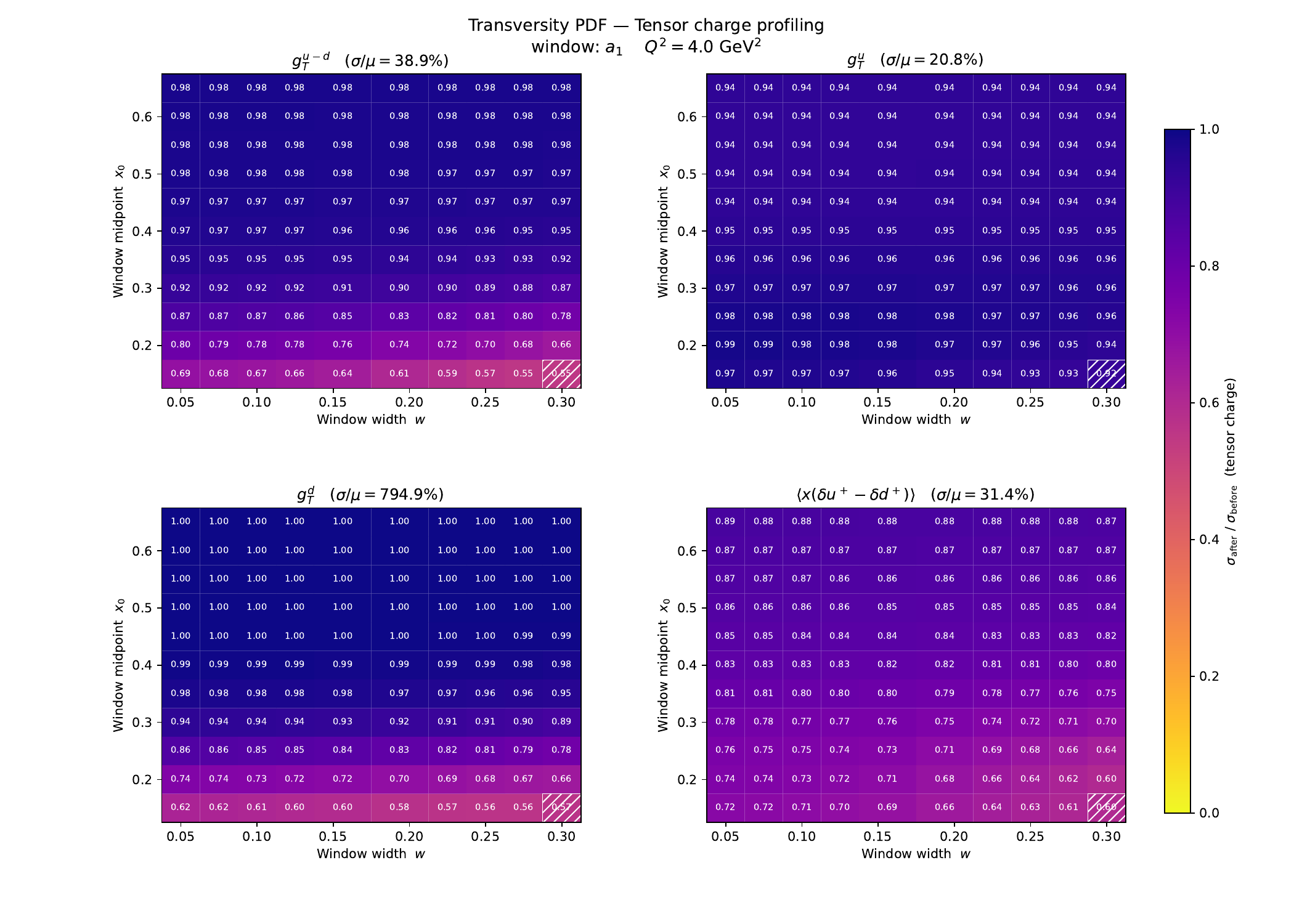}
  \caption{Higher-order moment Hessian update for $a_1(\xz,w)$ at $\varepsilon
  = 10\%$. Compare to figure~\ref{fig:charges_a0}}
  \label{fig:charges_a1}
\end{figure*}

We investigate the effect of updating on a higher-order window moment, $a_1$, in figure~\ref{fig:charges_a1}.
We see an overall reduction in constraining power when using a higher-order window moment. This follows from the fact that higher-order window moments are effectively narrower than the lower-order moments. We have already seen the width dependence of the constraining power in figure~\ref{fig:charges_a0}. The higher-order moment essentially shifts the heat maps left. 

In the next section we examine the effect of including multiple moments simultaneously in the profiling procedure.

\subsection{Accumulated moments}
\begin{figure*}[t]
    \centering
    \includegraphics[width=1\linewidth]{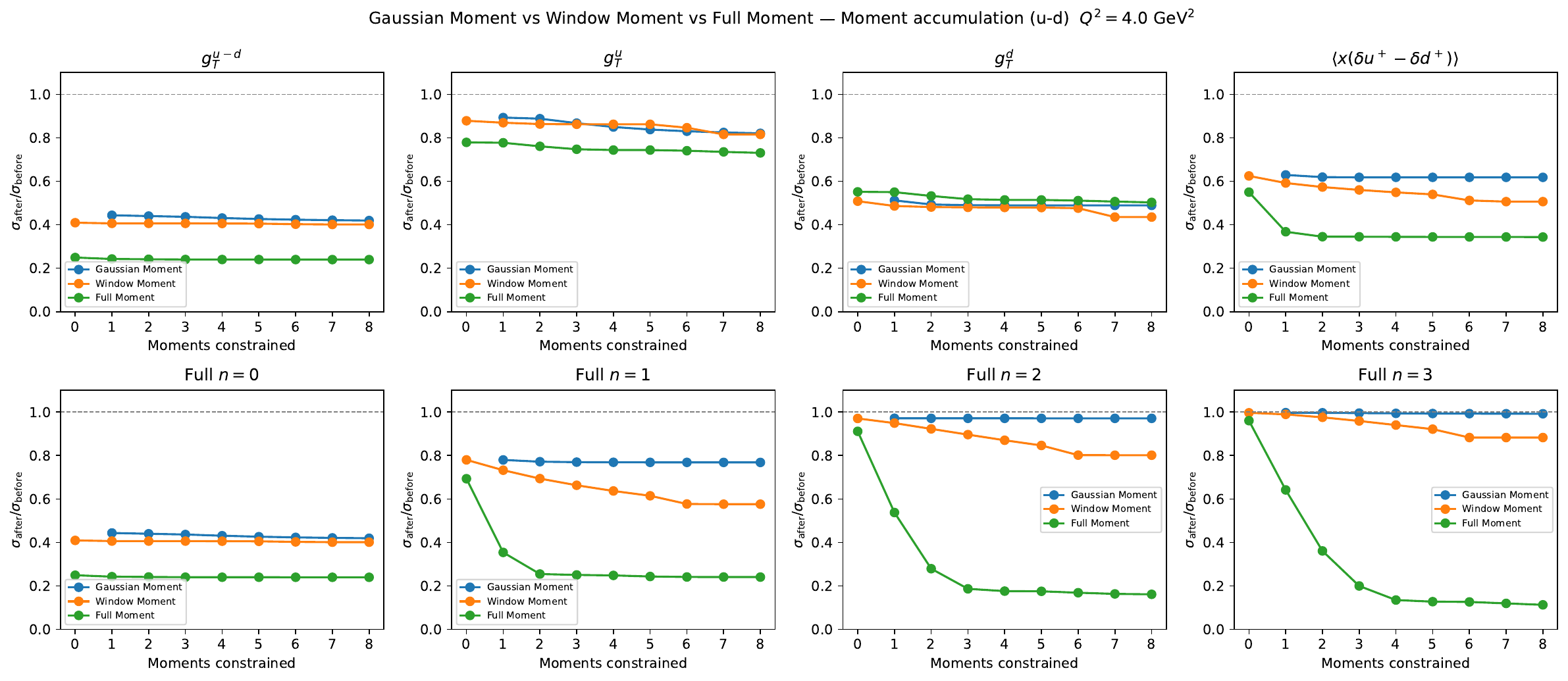}
    \caption{Comparison of the profiling of Gaussian, window, and full Mellin moments on the tensor charges and full moments. The Gaussian and window moments are at a midpoint, $x_0 = 0.15$, with width, $w=0.25$. Input pseudo data assume 10$\%$ relative uncertainty.}
    \label{fig:accumulated_moments}
\end{figure*}

In figure ~\ref{fig:accumulated_moments}, we show the comparison of the accumulated moments for the three kinds of input observable.
This figure was generated assuming all the input observables have a relative uncertainty of $10\%$. While this is not necessarily realistic, especially for higher moments, it allows for a more straightforward comparison of the window, Gaussian, and Mellin moments to determine, at a given relative uncertainty, what each lattice observables grants in terms of uncertainty reduction.

For simplicity, we also assumed no correlations between the input data for this update. This is an optimistic assumption that is reasonable for the full moments, which involve the computation of different matrix elements at each order (although some correlations will exist between calculations of different matrix elements on the same ensembles). This assumption fails for the window and Gaussian moments, but we anticipate that the inclusion of correlations will lead to relatively small effects. Indeed, the behavior of window and Gaussian moments is already quite flat with respect to the number of moments.

We see that, for $g_T^{u-d}$, the Mellin moments reduce the uncertainty on the tensor charge by $80\%$, compared to the window and Gaussian moments, which reduce it by $60\%$. For $g_T^{u}$ and $g_T^{d}$, the three approaches are much more similar. For the isovector momentum fraction, the Mellin moments demonstrate a clear advantage when both the zeroth and first moments are included.  

For the full moments, the full Mellin moments perform tautologically well when they are the input. Nevertheless, the results do show that the full moments, with the exception of the zeroth moment, are not well constrained by the window and Gaussian moments.

\begin{figure*}[t]
    \centering
    \includegraphics[width=1\linewidth]{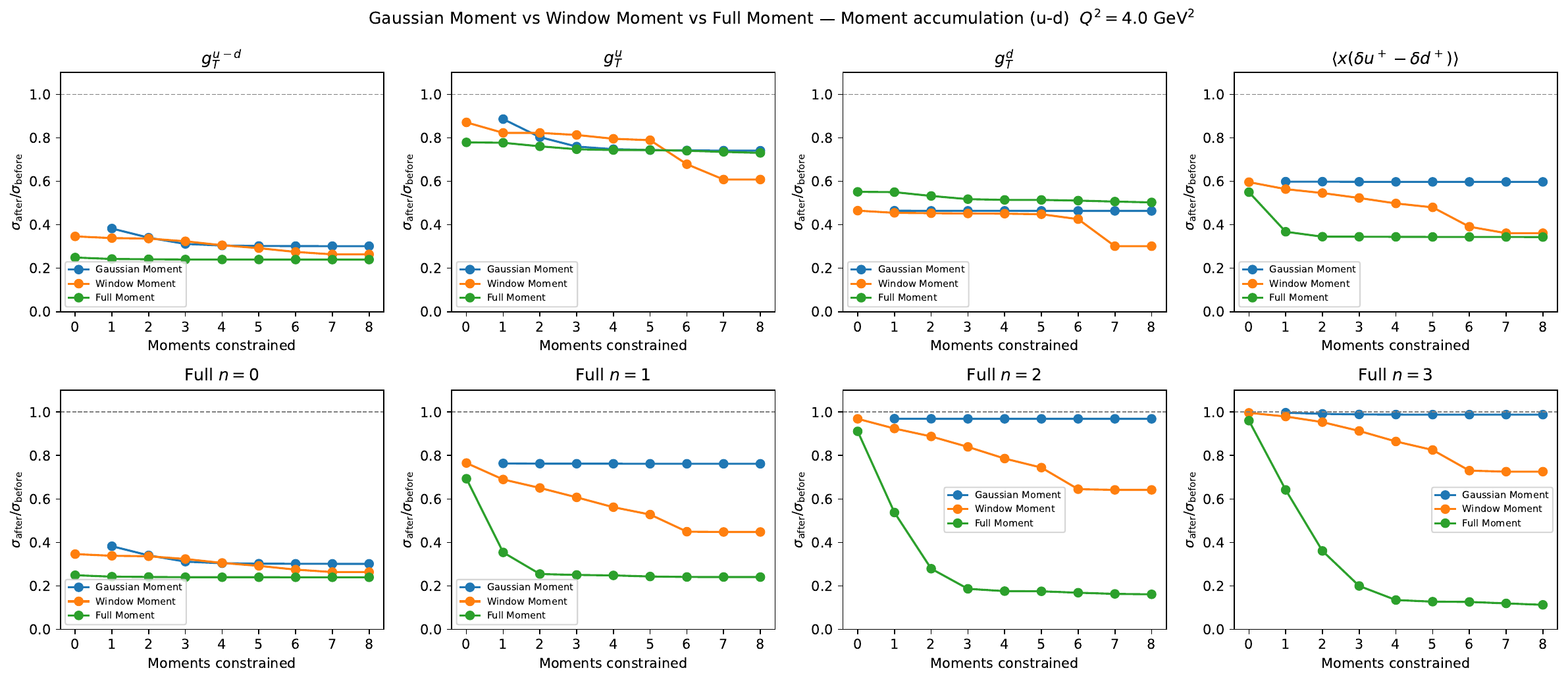}
    \caption{Same as figure~\ref{fig:accumulated_moments}, but with relative uncertainty of input pseudo data of 3$\%$.}
    \label{fig:accumulated_moments_precise}
\end{figure*}

In figure~\ref{fig:accumulated_moments_precise}, we repeat the exercise, but reduce the relative uncertainty of the input data to $3\%$.

Here we see a surprisingly strong dependence on the precision. For the tensor charge, $g_T^{u-d}$, the window and Gaussian moments are much closer in constraining power to the Mellin moments. The Mellin moment itself does not seem to improve in constraining power with higher precision, whereas the window moment improves with the inclusion of the sixth and seventh moments.

\subsection{Double scan}
\begin{figure*}[t]
    \centering
    \includegraphics[width=1\linewidth]{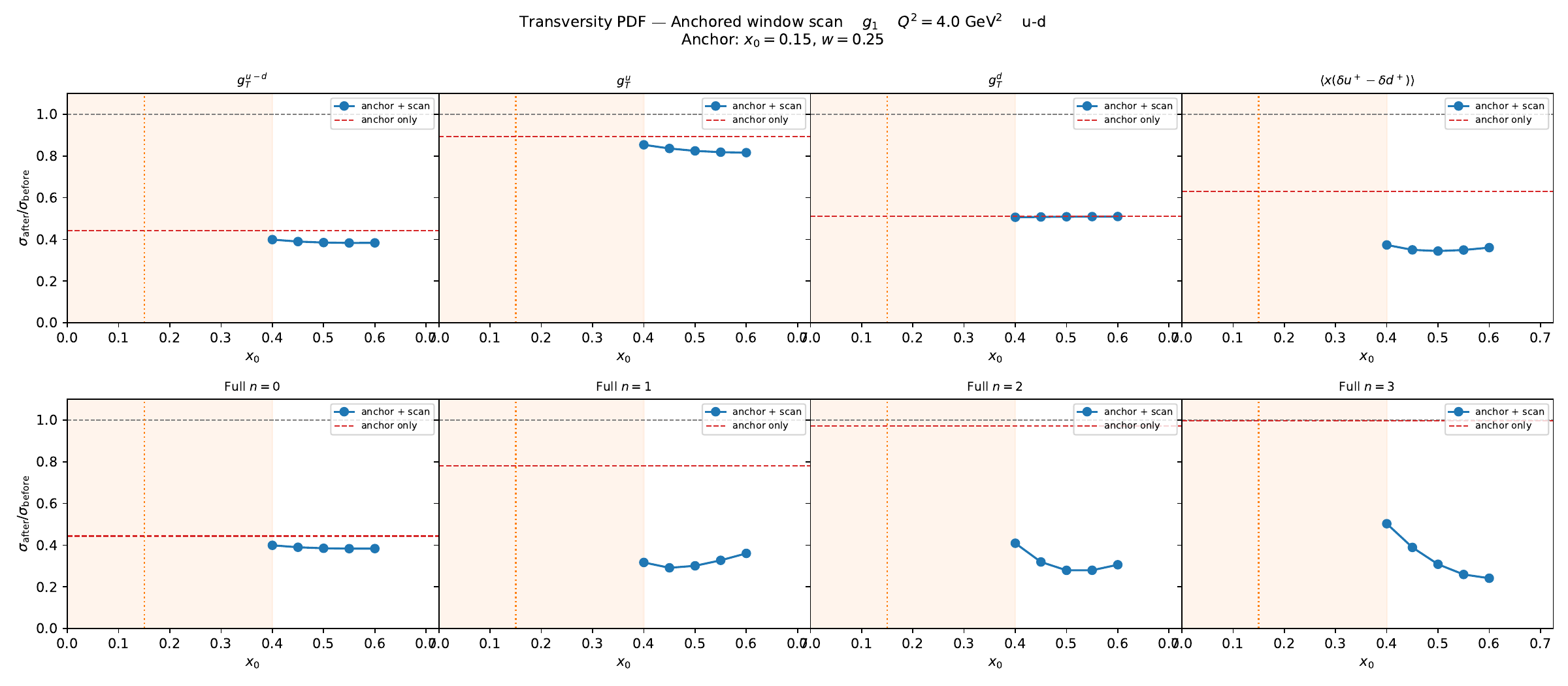}
    \caption{Improvement in tensor charge and full moment uncertainties when a single window moment is added, scanning over the midpoint of the new moment. The 'anchor' starting moment is at $x_0 = 0.15$, with width, $w=0.25$. Input pseudo data assume 10$\%$ relative uncertainty.}
    \label{fig:double_scan}
\end{figure*}

Lastly, after identifying the most promising window moment that constrains the tensor charges the best, at $x_0 = 0.15$, with width, $w=0.25$, we perform a second scan to identify where a second window moment should be computed to best constrain the uncertainty further. In order to avoid complications related to correlated uncertainties, we only consider windows that don't overlap with the first window. The second window is kept at the same width and relative uncertainty, $10\%$, as the first window. The results are shown in figure~\ref{fig:double_scan}.

We find improvements only in the isovector momentum fraction and the higher full moments. We also find that placing the second window further away from the first window tends to result in greater constraining power, but not monotonically so.

Other potential search algorithms that one could employ to find the best combination of two windows could perform more optimally than our 'greedy' algorithm, but we expect that figure~\ref{fig:double_scan} is nevertheless illustrative of the kind of improvements one can anticipate from combining two window moments.

\section{Summary}
\label{sec:results_summary}

PDFs capture the longitudinal momentum structure of the nucleon and are key inputs for precision studies at hadron colliders. Although PDFs have traditionally been extracted from global analyses of large ensembles of experimental data, recent developments in lattice QCD have enabled first-principles calculations of the $x$-dependence of PDFs and their Mellin moments up to $n=6$. Lattice results can provide independent and complementary information to global analysis or serve as inputs to PDF fits that are poorly constrained by existing experimental data. The transversity quark PDF, which represents the distribution of transversely polarized quarks in a transversely polarized nucleon, is the least known of the three leading-twist collinear
distributions and thus provides an ideal quantity to examine the impact of lattice results on global analyses.

Moreover, the emerging field of $x-$dependent hadron structure calculations on the lattice necessitates the development of quantitative precision benchmarks and cross-checks between lattice calculations and between lattice and phenomenological results. The authors of Ref.~\cite{Karpie:2025ikc} proposed window observables as one such set of benchmark quantities, inspired by the success of similar approaches to benchmarking lattice calculations of the hadronic vacuum polarization contribution to the muon anomalous magnetic moment. Window observables exploit the non-trivial point-by-point correlations in the $x$-dependent results extracted from lattice data, which ensure that integrals over $x$ can be determined far more precisely than the point-wise uncertainty might suggest. 

In this work, we quantify the impact that lattice QCD window observables may have on PDF uncertainties if incorporated into a global analysis, using the \epump{} framework. This framework uses the Hessian approximation to estimate the impact of new data on an existing PDF set and its uncertainties at a fraction of the
computational cost of a full refit. In particular, we study the impact of window observables and Mellin moments on the isovector transversity quark PDF. We profile over a range of windows of different midpoints and widths and explore the impact
on the transversity tensor charges and isovector momentum fraction. We also study the ability of the window and Gaussian moments to constrain the full Mellin moments and compare results to a profiling of the full Mellin moments to ascertain the uncertainty reduction trade-off between window, Gaussian and full Mellin moments.

We find that the windows and Gaussian moments provide
largely equivalent constraints on the quark transversity PDF
and tensor charge, given similar precision for both windows observables. This result is not surprising, because both the window moment and the Gaussian window carry similar nonperturbative information from any given lattice calculation. The caveat here, relevant for future studies, is that it may be possible to construct Gaussian windows with higher precision than window moments, given the same lattice data, because the Gaussian window can be constructed to weight more highly the most precise region in momentum fraction or Ioffe time.

We demonstrate that the window midpoint $\xz$ is the primary determinant of
constraining power, with window midpoints in the range $\xz \approx 0.15$--$0.25$
give the best reduction on $g_T^{u-d}$. This midpoint is slightly lower than the value of  $\xz = 0.3$ used in Ref.~\cite{Karpie:2025ikc}. In contrast, the effect of the window width $w$ is modest for the range of values we studied. 

Importantly for phenomenology, our results show that a single isovector window measurement at $\varepsilon = 10\%$ can reduce the uncertainty on $g_T^{u-d}$ from around 39\% to roughly 11\% (a factor of $\approx 3.6$) at the optimal window placement. The identity $g_T^d = g_T^u - g_T^{u-d}$ and the
near-zero correlation between $g_T^u$ and the window observable propagates this improvement almost entirely to $g_T^d$, reducing a prior
uncertainty of 795\% by approximately a factor of 3.6. The $u$-quark
tensor charge $g_T^u$ remains essentially unaffected by the window moment constraint. Future lattice calculations that control all sources of uncertainty will lead to improved window observables. Our analysis shows that $\varepsilon = 10\%$ uncertainty roughly doubles the uncertainty reduction on the tensor charge obtained from global analysis, compared to an uncertainty of $\varepsilon = 25\%$.

We compare the constraining power between the full Mellin moments and the window moments, and how this constraining power behaves as a function of the number of moments. We see that, at a given precision level, the Mellin moments have greater constraining power than the window moments, but the window moments benefit more from added precision and moment orders. Thus, it remains important for the lattice community to pursue extractions of $x$-dependent distributions themselves in conjunction with calculations of higher Mellin moments directly from local operators. 

Last, we study the effect of including two window moments at different midpoints. This does not have a large impact on the tensor charge, $g_T^{u-d}$, but does have a large effect on the isovector momentum fraction and the higher order moments.

The use of multiple strategies with the lattice community has had significant benefits in the flavor sector~\cite{FlavourLatticeAveragingGroupFLAG:2024oxs} and for the anomalous magnetic moment of the muon~\cite{Aliberti:2025beg}. Our results demonstrate that it is likely to continue to have a greater impact on our understanding of hadron structure than narrowing the focus on just one or two methods or observables. In particular, the impact of window observables or Mellin moments on global analysis is just one application for these quantities. Both provide benchmark quantities that enable quantitative comparison between lattice calculations and existing phenomenological fits and, in particular, window observables offer advantages as benchmarks and cross-checks for $x$-dependent lattice frameworks and methods .

\section{Acknowledgments}
We thank Peter Risse for the suggestion to use \epump{} to explore the impact of lattice observales.
We thank members of the HadStruc Collaboration for helpful discussions of window moments in lattice QCD. C.J.M.~is supported in part by U.S.~DOE Grant \mbox{\#DE-SC0025908}. DA is grateful for support from the Wu-Ki Tung Endowed Chair in particle physics. This work has benefited from the collaboration enabled by the Quark-Gluon Tomography (QGT) Topical Collaboration, U.S.~DOE Award \mbox{\#DE-SC0023646}.

During the preparation of this manuscript, DA used Claude Sonnet 4.6 (Anthropic) to brainstorm ideas, to create a computational  pipeline that connects  \epump{}, \mchess{}, and code written by the authors, generate figures, and provide an initial outline of the paper. The authors wrote, edited, and approved all final text and take full responsibility for the content.

\appendix
\section{Supplementary heatmaps}
\label{app:heatmaps}

This appendix collects the complete set of \epump{} profiling heatmaps
for all input observables and output quantities studied. The main text
discusses the $a_0$ and $g_1$ charge comparisons
(Figs.~\ref{fig:charges_a0},\ref{fig:charges_g1}) and the $\varepsilon = 25\%$ precision
comparison (Fig.~\ref{fig:precision_comparison}) in detail. The figures
below provide the complete picture.

\begin{figure}[ht]
  \centering
  \includegraphics[width=0.5\textwidth]{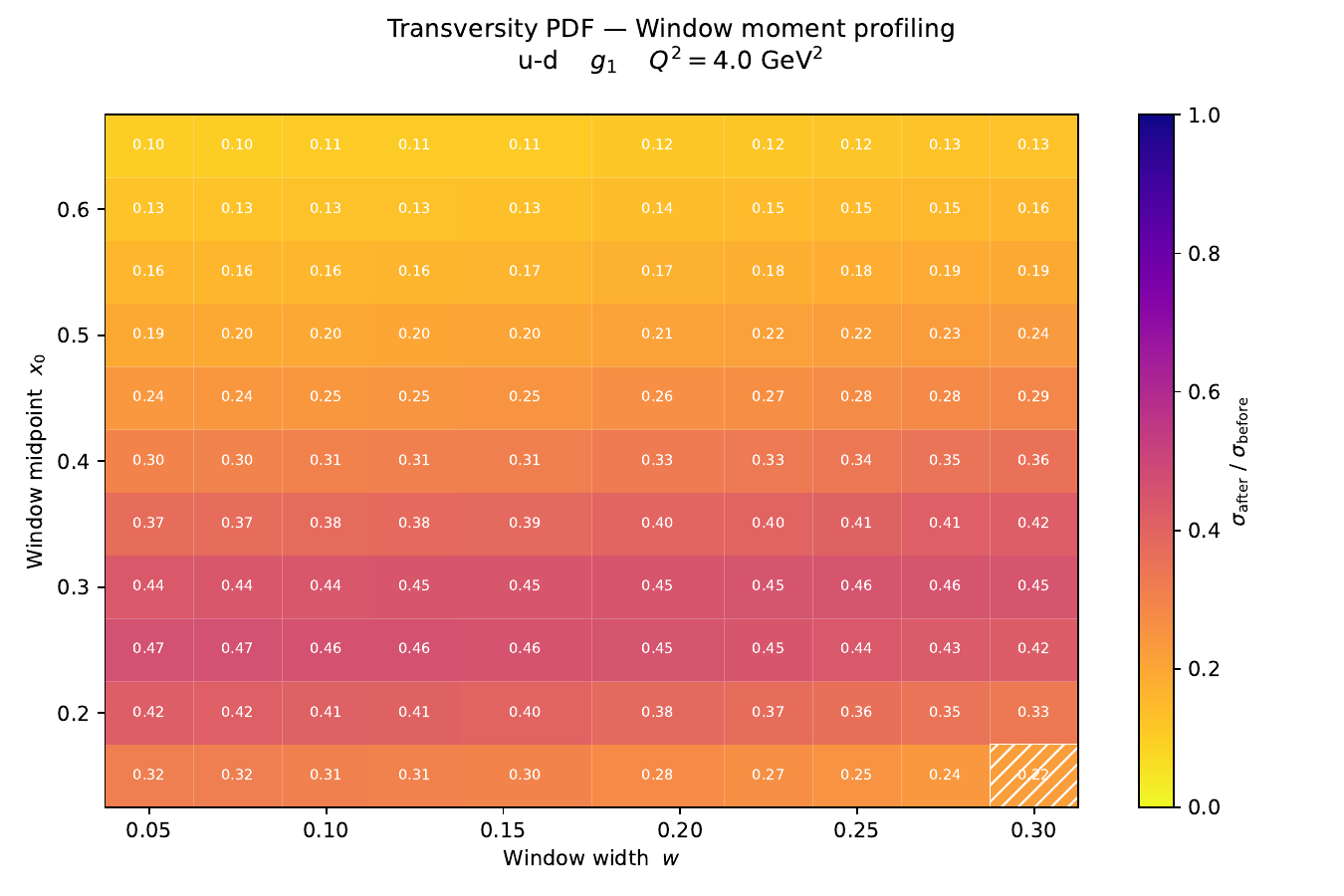}
  \caption{Input observable cross-check for $g_1(\xz,w)$ at $\varepsilon
  = 10\%$. The result is fully predicted by simple error propagation
  (Eq.~\protect\eqref{eq:harmonic}) and confirms correct pipeline
  operation. The variation across the plane maps the relative PDF
  uncertainty on $g_1$. JAMDiFF23, $Q^2 = 4.0\,\text{GeV}^2$.}
  \label{fig:self_g1}
\end{figure}

\begin{figure}[ht]
  \centering
  \includegraphics[width=0.5\textwidth]{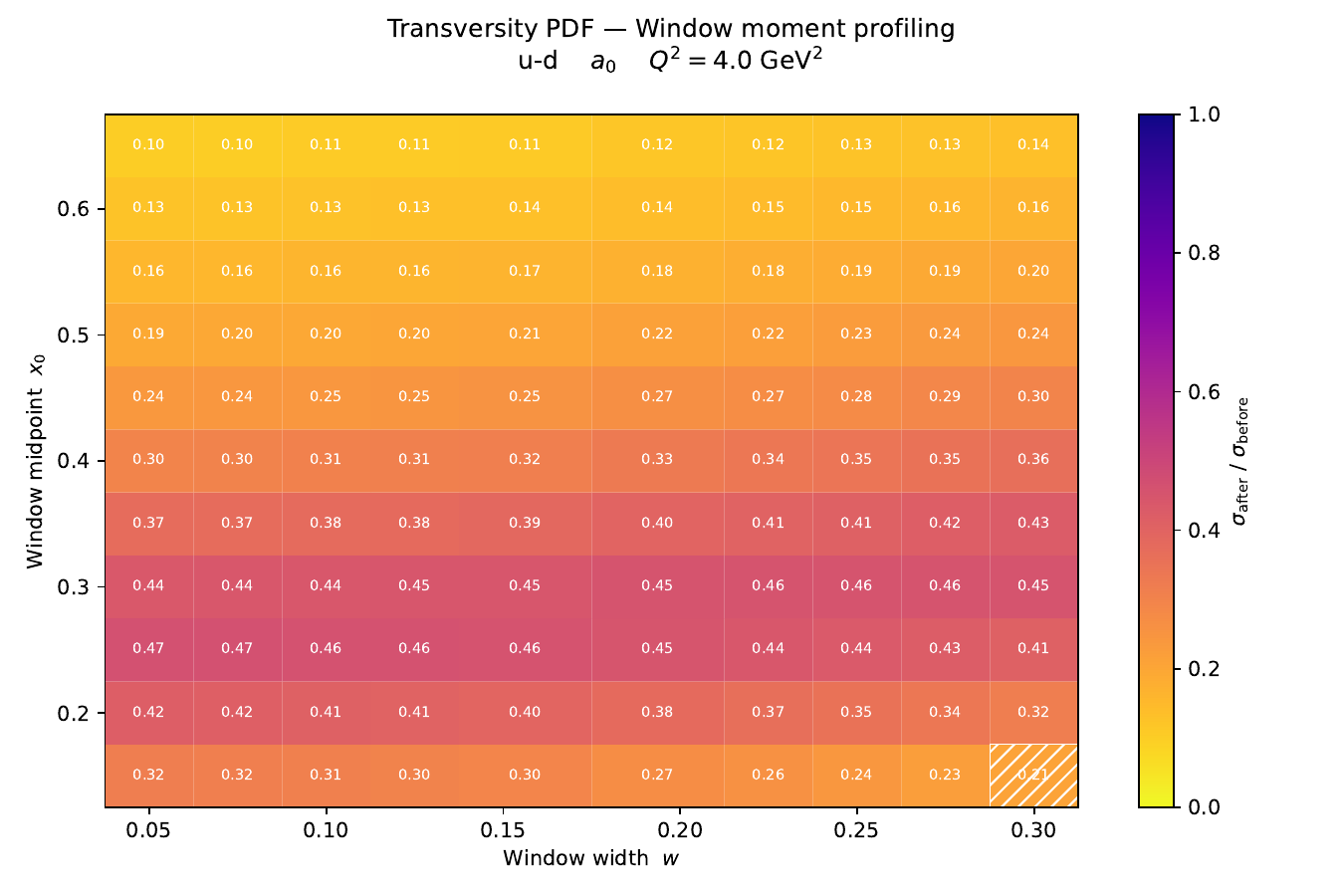}
  \caption{Input observable cross-check for $a_0(\xz,w)$ at $\varepsilon
  = 10\%$. Cf.\ Fig.~\ref{fig:self_g1}. The two families yield nearly
  identical patterns, confirming their equivalent sensitivity to the PDF
  eigenvector structure. JAMDiFF23, $Q^2 = 4.0\,\text{GeV}^2$.}
  \label{fig:self_a0}
\end{figure}

\begin{figure}[ht]
  \centering
  \includegraphics[width=0.5\textwidth]{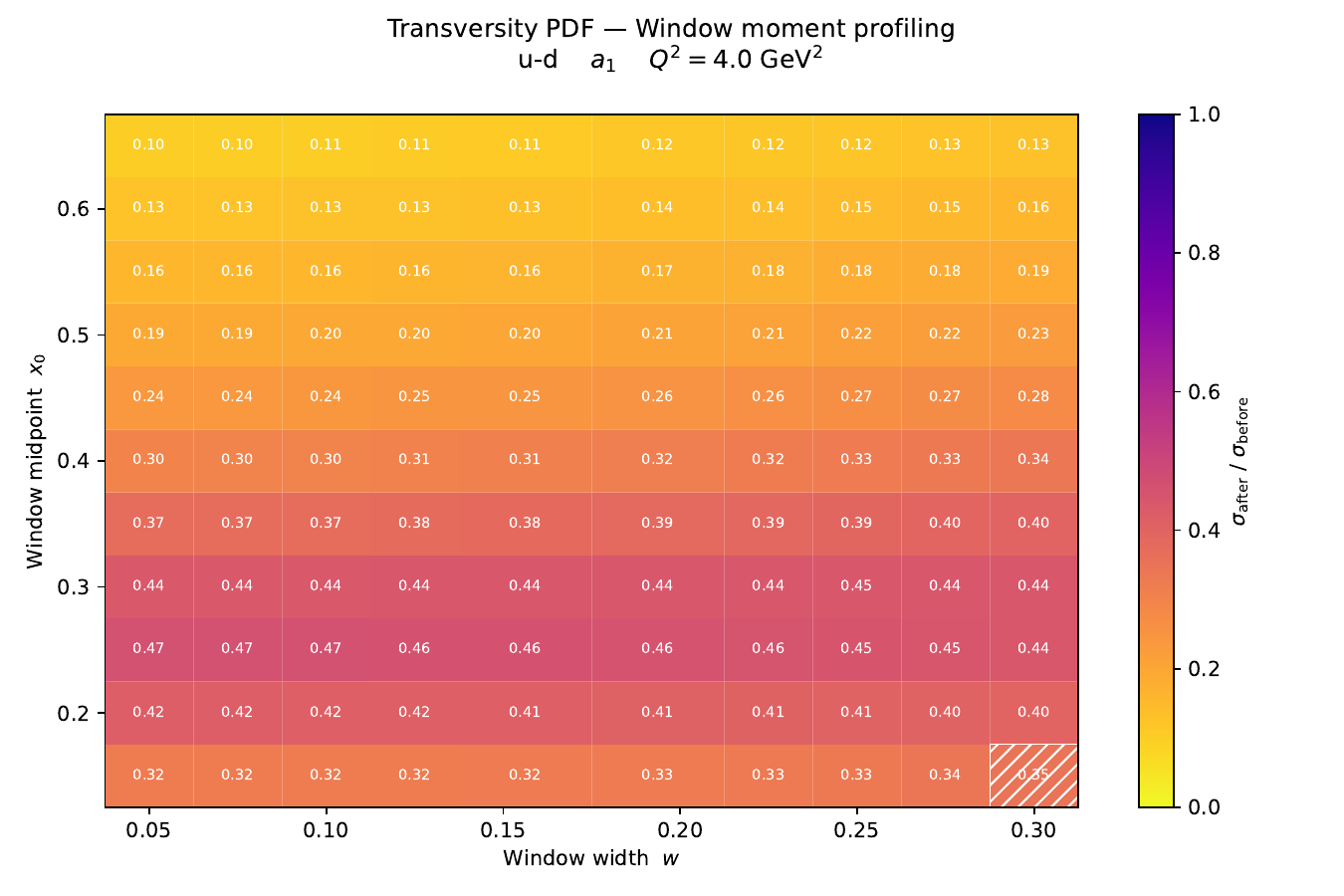}
  \caption{Input observable cross-check for $a_1(\xz,w)$ at $\varepsilon
  = 10\%$. The $n{=}1$ flat moment shows greater width-independence than
  $a_0$ or $g_1$, reflecting the $x$-weighted integrand's reduced
  sensitivity to the low-$x$ region where the PDF varies most rapidly.
  JAMDiFF23, $Q^2 = 4.0\,\text{GeV}^2$.}
  \label{fig:self_a1}
\end{figure}

Figures~\ref{fig:self_g1}--\ref{fig:self_a1} show
$\sigma_\text{after}/\sigma_\text{before}$ on the input window observable
itself, which is the trivial $\cos\theta = 1$ case fully predicted by
Eq.~\eqref{eq:harmonic}. The structure across the $(\xz,w)$ plane is
simply a map of the relative PDF uncertainty $\Delta X / X$ on the window
observable. Values approaching zero at large $\xz$ reflect regions where
the quark transversity PDF is very small and the PDF uncertainty is large
relative to the assumed $\varepsilon$; values near unity at small $\xz$
indicate regions where the two precisions are comparable. These plots
confirm that the \epump{} pipeline is operating correctly and motivate
the choice of $\varepsilon = 10\%$ as the threshold below which meaningful
tensor charge constraints are achievable.

\begin{figure*}[h]
  \centering
  \includegraphics[width=\textwidth]{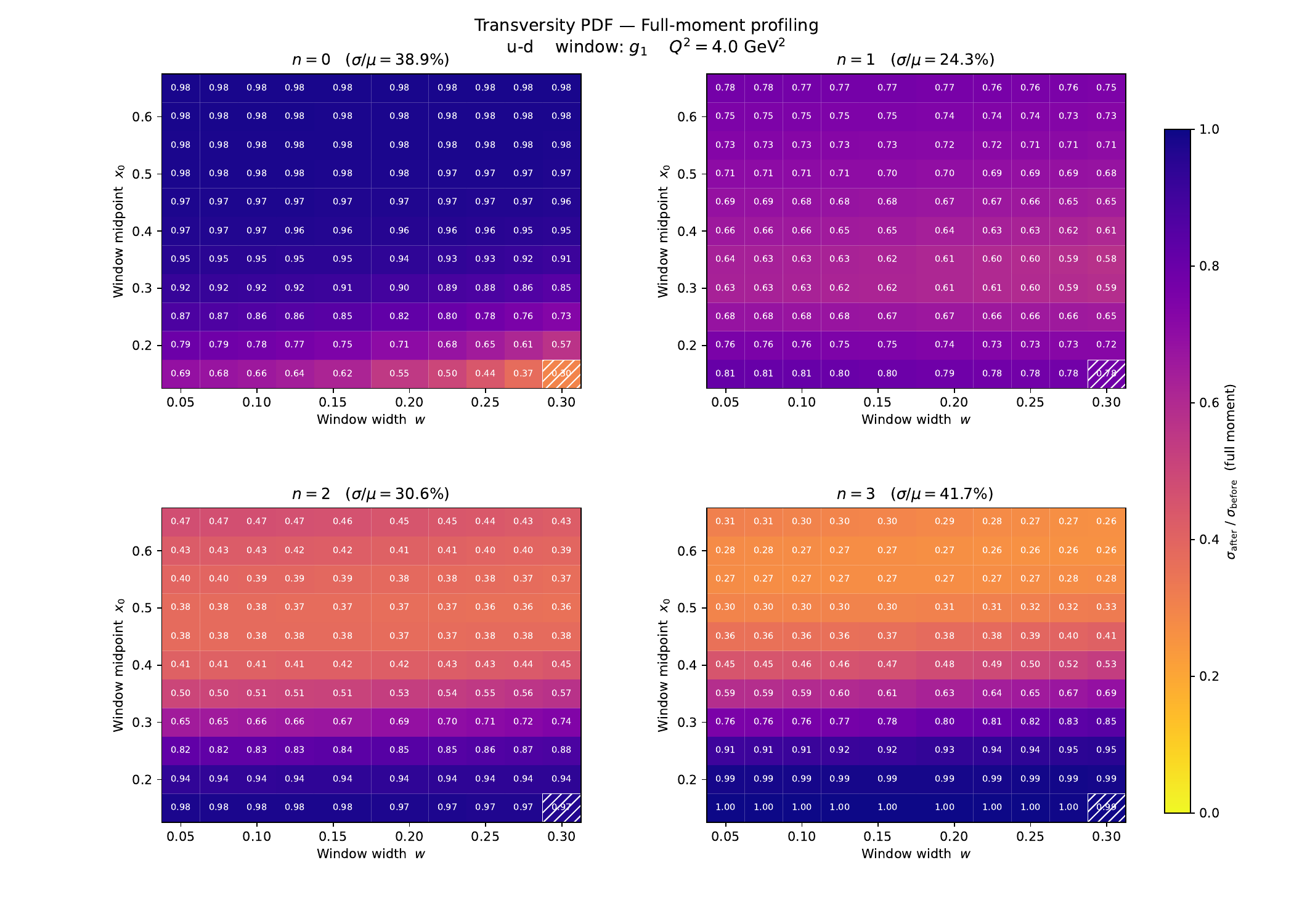}
  \caption{$\sigma_\text{after}/\sigma_\text{before}$ on the full Mellin
  moments $a_n(0,1)$ for $n = 0,1,2,3$ after profiling with $g_1(\xz,w)$
  at $\varepsilon = 10\%$. The $n{=}2$ panel shows the weakest reduction
  (smallest $\cos\theta$ between $g_1$ and $x^2$-weighted integrand);
  the $n{=}3$ panel shows the strongest reduction at high $\xz$ due to
  its large prior uncertainty amplifying any non-zero correlation.
  JAMDiFF23, $Q^2 = 4.0\,\text{GeV}^2$.}
  \label{fig:moments_g1_10}
\end{figure*}

\begin{figure*}[h]
  \centering
  \includegraphics[width=\textwidth]{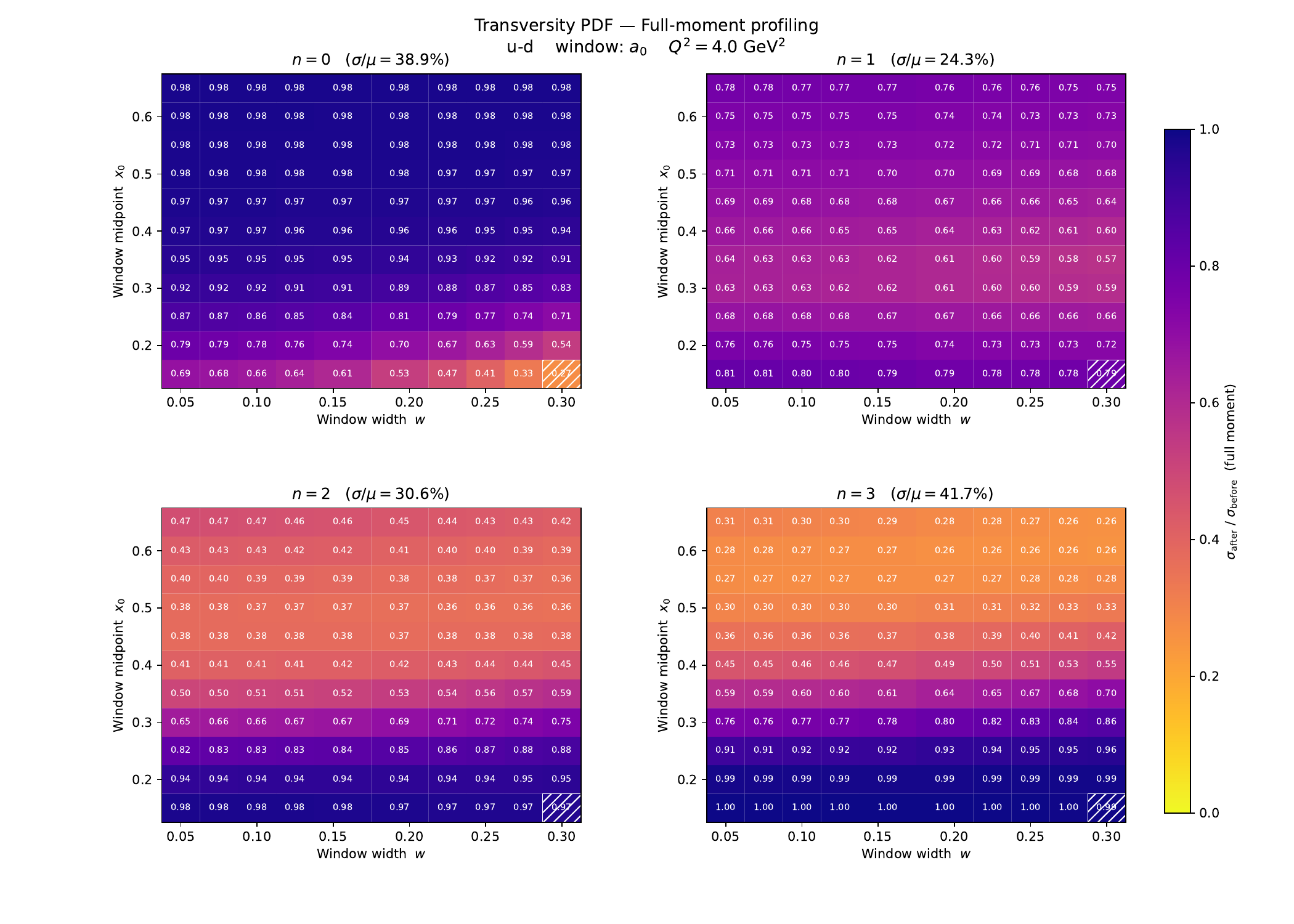}
  \caption{As Fig.~\ref{fig:moments_g1_10} but with $a_0(\xz,w)$ as
  input. The four panels are nearly indistinguishable from those of
  Fig.~\ref{fig:moments_g1_10}, confirming the equivalence of the two
  window families at fixed precision.
  JAMDiFF23, $Q^2 = 4.0\,\text{GeV}^2$.}
  \label{fig:moments_a0_10}
\end{figure*}

\begin{figure*}[h]
  \centering
  \includegraphics[width=\textwidth]{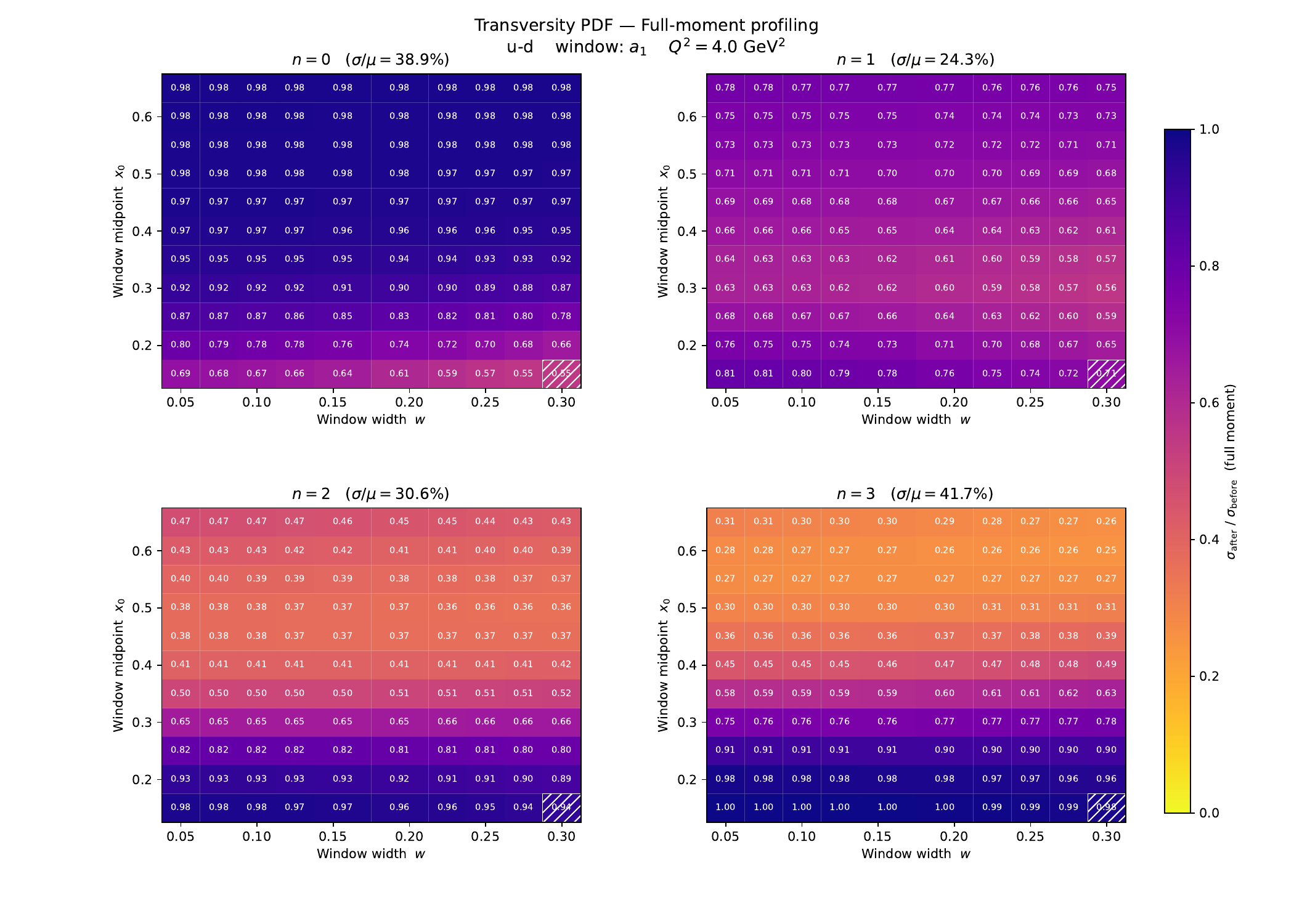}
  \caption{As Fig.~\ref{fig:moments_g1_10} but with $a_1(\xz,w)$ as
  input. The $x$-weighted flat moment shows slightly weaker reduction
  on $g_T^{u-d}$ ($n{=}0$) but comparable reduction on higher-$n$
  moments, reflecting the better overlap of its integrand with
  $x^n$-weighted outputs at $n \geq 1$.
  JAMDiFF23, $Q^2 = 4.0\,\text{GeV}^2$.}
  \label{fig:moments_a1_10}
\end{figure*}

Figures~\ref{fig:moments_g1_10}--\ref{fig:moments_a1_10} show
$\sigma_\text{after}/\sigma_\text{before}$ on the full Mellin moments
$a_n(0,1)$ for $n = 0,1,2,3$ for each input observable at $\varepsilon
= 10\%$.

\begin{figure*}[h]
  \centering
  \includegraphics[width=\textwidth]{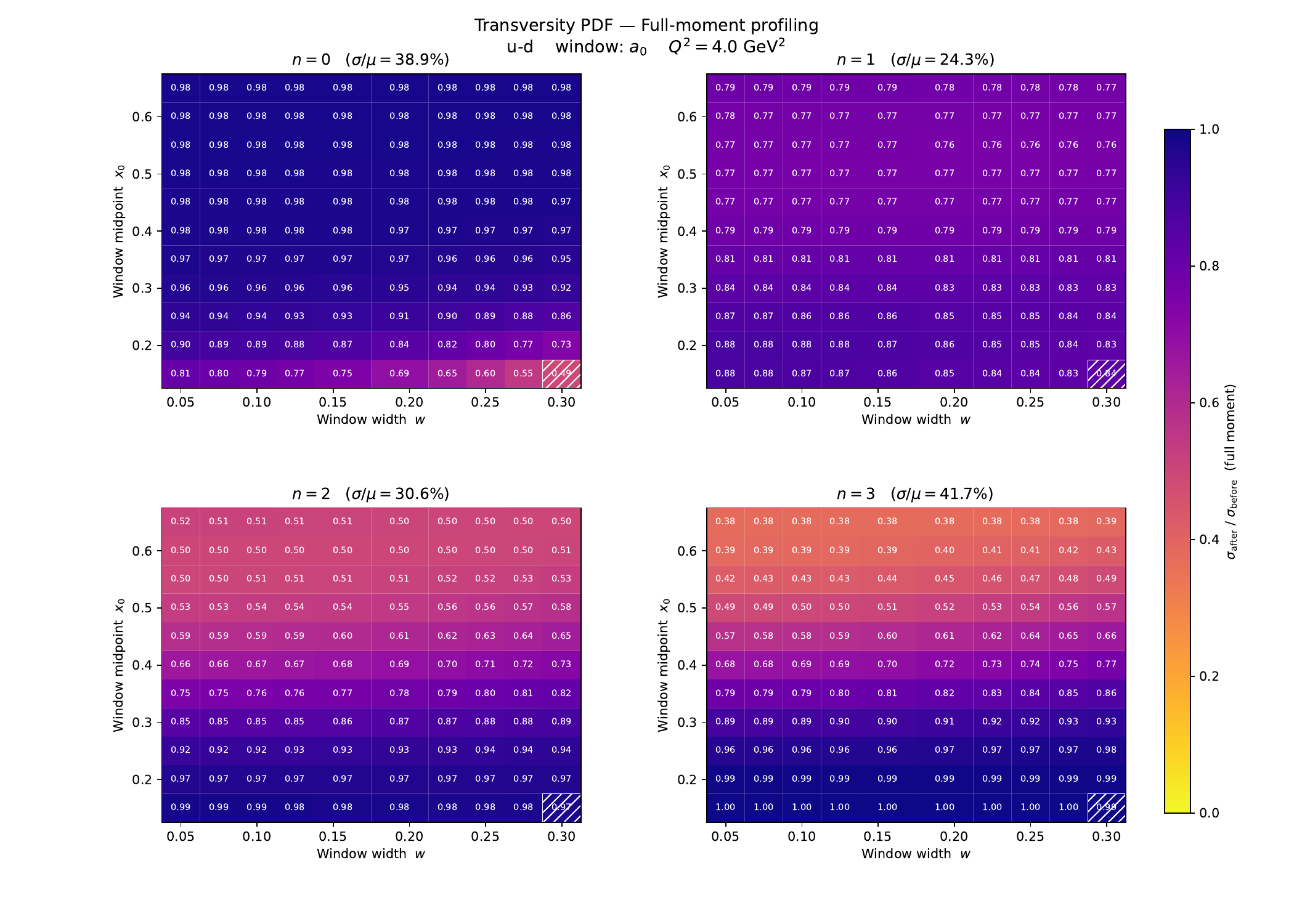}
  \caption{As Fig.~\ref{fig:moments_a0_10} but with a $20\%$ relative uncertainty.
  JAMDiFF23, $Q^2 = 4.0\,\text{GeV}^2$.}
  \label{fig:moments_a0_25}
\end{figure*}

Figures~\ref{fig:moments_a0_25} full moment profiling with $a_0$ input at $\varepsilon = 25\%$
respectively. Together with Fig.~\ref{fig:moments_a0_10} and Fig.~\ref{fig:moments_g1_10} in the main
text, these demonstrate the robustness of the main conclusions across
input observable families and lattice precision assumptions.

\bibliographystyle{apsrev4-2}

\end{document}